\documentclass[doublespace,AMA,STIX1COL]{WileyNJD-v2}

\articletype{Article Type}%

\received{9 August 2022}
\revised{<Date> <Month> <Year>}
\accepted{<Date> <Month> <Year>}

\begin{document}

\title{Determining Parameter Ranges for High Accuracy Large Eddy Simulation by Lax-Wendroff Method\protect\thanks{Parameters for High Accuracy LES for Lax-Wendroff Method}}

\author[1,2]{V.K. Suman*}

\author[3]{Soumyo Sengupta}

\author[2]{P. Sundaram}

\author[4]{Aditi Sengupta}

\author[4]{Tapan K. Sengupta}

\authormark{V.K. SUMAN \textsc{et al}}

\address[1]{\orgdiv{Computational \& Theoretical Fluid Dynamics Division}, \orgname{CSIR-NAL}, \orgaddress{\state{Karnataka}, \country{India}}}

\address[2]{\orgdiv{High Performance Computing Laboratory, Aerospace Engineering}, \orgname{Indian Institute of Technology Kanpur}, \orgaddress{\state{Uttar Pradesh}, \country{India}}}

\address[3]{\orgdiv{CERFACS}, \orgname{42 Avenue G. Coriolis, 31057}, \orgaddress{\state{Toulouse Cedex 1}, \country{France}}}

\address[4]{\orgdiv{Dept. of Mechanical Engineering}, \orgname{Indian Institute of Technology (ISM) Dhanbad}, \orgaddress{\state{Jharkand}, \country{India}}}

\corres{*V.K. Suman, CTFD Division, CSIR-NAL. \email{vksuman@iitk.ac.in}}

\presentaddress{CTFD Division, CSIR-NAL, India.}

\abstract[Abstract]{The analysis of Lax-Wendroff (LW) method is performed by the generic modified differential equation (MDE) approach in the spectral plane using Fourier transform. In this approach, the concept of dispersion relation plays a major role relating spatial and temporal dependence of the governing differential equation, including initial and boundary conditions in developing high accuracy schemes. Such dispersion relation preserving schemes are calibrated in the spectral plane using the global spectral analysis for the numerical method in the full domain. In this framework, the numerical methods are calibrated by studying convection and diffusion as the underlying physical processes for this canonical model problem. In the LW method spatial and temporal discretizations are considered together, with time derivatives replaced by corresponding spatial derivatives using the governing equation. Here the LW method is studied for the convection-diffusion equation (CDE) to establish limits for numerical parameters for an explicit central difference scheme that invokes third and fourth spatial derivatives in the MDE, in its general form. Thus, for the LW method, two different MDEs are obtained, depending on whether the LW method is applied only on the convection operator, or both on the convection and diffusion operators. Motivated by a one-to-one correspondence of the Navier-Stokes equation with the linear CDE established in ``Effects of numerical anti-diffusion in closed unsteady flows governed by two-dimensional Navier-Stokes equation- Suman et al. Comput. Fluids, 201, 104479 (2020)'', an assessment is made here to solve flow problems by these two variants of the LW method. Apart from mapping the numerical properties for performing large eddy simulation for the LW methods, simulations of the canonical lid-driven cavity problem are performed for a super-critical Reynolds number for a uniform grid.}



\maketitle

\footnotetext{\textbf{Abbreviations:} CDE, convection-diffusion equation; GSA, global spectral analysis; NSE, Navier-Stokes equation; LW, Lax-Wendroff; MDE, modified differential equation}

\section{Introduction}
Progress in scientific computing has been made possible due to the developments in hardware and analysis of numerical methods. Classical analysis use the order of the numerical methods based on Taylor's series expansion for the spatial and temporal discretization adopted in solving the governing partial differential equation. Different governing equations require {\it a priori} analysis for suitability of a numerical method. For example, using uniformly second order central differencing scheme for space and time derivatives led to a solution for the canonical one-dimensional (1D) convection equation,

\begin{equation}
\frac{\partial u}{\partial t} + c\frac{\partial u}{\partial x} = 0
\label{eqn1}
\end{equation}

\noindent with the solution remaining bounded strictly \cite{CFL} for this mid-point leap frog method \cite{WFAmes}. But, for the canonical heat equation,

\begin{equation}
\frac{\partial u}{\partial t} = \alpha \frac{\partial^2 u}{\partial x^2}
\label{eqn2}
\end{equation}

the numerical method becomes unconditionally unstable for the same second order discretization \cite{WFAmes,HACM} for this parabolic partial differential equation, with the method attributed to Richardson \cite{LFR22}. The hyperbolic partial differential equation \eqref{eqn1} with the midpoint leap frog method played a major role in earlier weather prediction.

The Fourier series analysis of heat equation by von Neumann and Richtmyer \cite{LA657,HACM41,Morton_Mayers} explained why the second order discretization of heat equation is unconditionally unstable, which is due to the creation of a spurious unstable numerical mode. Despite the early success of this numerical stability study, it is noted \cite{HACM} that Fourier series analysis has some unresolved issues that prompted Zingg \cite{HACM362} to state that {\it through Fourier analysis, one can evaluate the phase and amplitude error of a given method as a function of wavenumber. However, this information can be difficult to interpret.} The author is referring to the inability of Fourier series analysis method to explain the correct dispersion error of any numerical method for space-time dependent governing equation.

The main limitation of Fourier analysis is due to its strict validity for spatially periodic problem only, for the governing differential equations with constant coefficient. Also one performs a normal mode analysis, where modes do not interact with each other, even for a linear governing equation. Thus, despite the Fourier analysis method being easy to apply, it is not considered further due to its inability to solve general non-periodic problems while including effects of boundary conditions.

Analysis of the heat equation (Eq. \eqref{eqn2}) is mainly used for numerical stability at the steady state, without considering about the transient state. This is not like the convection equation (Eq. \eqref{eqn1}), for which the numerical solution must propagate at the correct speed, without dissipation and dispersion. Thus, this is a more relevant canonical problem for numerical analysis, as noted in \cite{CFL}. The main application for this equation was for numerical weather prediction \cite{Haltiner_Williams}, where tracking the propagating disturbances (as resolved into acoustic, vortical and entropic waves) are vitally important \cite{HACM,Haltiner_Williams,Durran,HACM363}. Relevance of this equation is explained briefly in terms of the concept of dispersion relation preservation (DRP) property in the following section.

The numerical analysis of Eq. \eqref{eqn1} involves the dependence of the unknown on space and time simultaneously, and one writes the general representation of the unknown in terms of Fourier-Laplace transform as,

\begin{equation}
u(x,t) = \int \int \hat{U}(k, \omega_0) e^{i(kx - \omega_0 t)} dk\; d\omega_0
\label{eqn3}
\end{equation}

Substituting Eq. \eqref{eqn3} in \eqref{eqn1}, one gets the physical dispersion relation as,

\begin{equation}
\omega_0 = kc
\label{eqn4}
\end{equation}

This relates the spatial and temporal scales in the spectral plane. For the computation of Eq. \eqref{eqn1}, one notes the spatial and temporal discretizations to be such that Eq. \eqref{eqn4} is satisfied for every wavenumber, with a corresponding circular frequency for the DRP schemes. For such schemes, one can show that the energy of the system travels at the group velocity defined \cite{HACM} by,
$$v_g = \frac{d\omega_0}{dk}$$

Comparing Eq. \eqref{eqn1} with Eq. \eqref{eqn4}, it is noted that $v_g = c$. One can also show that if the initial condition to solve Eq. \eqref{eqn1} is given by, $u_0 (x) = \int \hat{U}_0 (k) e^{ikx} dk$, then the general solution at a later time is given by, $u(x,t) = u_0 (x -ct)$. This implies that the initial amplitude remains unchanged, as the solution propagates downstream with the speed $c$, for all the spatial scales associated with the initial condition. Thus, the solution of Eq. \eqref{eqn1} is non-dissipative and non-dispersive. One notes that $u(x,t)$ in Eq. \eqref{eqn3} via the integration of the variable (along its strip of convergence) ensures all modal and nonmodal interactions. This is the basis of the global spectral analysis (GSA), which is distinct from the von Neumann analysis. GSA has been advocated earlier in \cite{HACM363,HACM114} and comprehensive details can be noted in \cite{HACM}.

In \cite{HACM,Haltiner_Williams}, the major achievement was in identifying the correct numerical dispersion relation. The authors underscored the importance of spatial and temporal discretizations to obtain a numerical phase speed which is different from the physical phase speed, such that the numerical dispersion relation is given by,

\begin{equation}
\omega_{num} = kc_{num}
\label{eqn5}
\end{equation}

Note that in GSA, $x$ or $k$ ($x$ is the space variable and $k$ is wavenumber) is the independent variable, while the temporal scale is related to $k$ via the numerical dispersion relation. A DRP method is one, for which the departure of the numerical dispersion relation from the physical dispersion relation is negligible for accessible ranges of numerical parameters. The quantitative analysis used in GSA \cite{HACM,HACM256,HACM259} is based on the modified differential equation (MDE) approach that is described next.

Notable examples of the MDE are for (i) more accurate numerical methods for parabolic equations in Milne's method \cite{HACM175}, 
and (ii) for `improved' numerical stability in Dufort-Frankel's method \cite{HACM65}. It is known that increased stability for Dufort-Frankel's method \cite{HACM65} also brings in the problem of inconsistency \cite{WFAmes,HACM} for some numerical parameter combinations. In the MDE, one reconverts the discretized equation to an equivalent differential equation, as shown in \cite{Warming_Hyett}. Linear problems with variable coefficients have been studied in \cite{Garabedian} and nonlinear problems are shown in \cite{Harten_etal} by the MDE. However, Li and Yang \cite{Li_Yang} made a generalization that the MDE is {\it very heuristic, unfortunately just valid for solutions in smooth regions or at low frequency modes \cite{Warming_Hyett}: Therefore the connection with the von Neumann analysis is only restricted there}, and cited \cite{Griffiths} that there is a lack of theoretical foundation for the MDE.

The MDE continues to be used in Lax-Wendroff (LW) method for hyperbolic partial differential equations \cite{HACM151,Winnicki}. Apart from these, one notes that the Russian school \cite{HACM286,Yanenko} explained different aspects of the MDE. Acoording to their classification, if one converts the difference equation into an equivalent differential equation form by retaining both space and time derivatives, then it is called the $\Gamma$- form analysis. On the other hand, if the discrete equation is converted back in the differential form with all the truncation terms converted in terms of the spatial derivatives, then that is termed as the $\Pi$-form analysis. Variants of the $\Gamma$- and $\Pi$-forms of the MDE along with Fourier-Laplace representation of the unknowns are used in development of the GSA and explained in the following.

The LW method is one of the popular methods for higher order solution of PDEs and continues to be used in the CFD community \cite{Wang_etal,Brunet_etal,Schoenfeld_Rudgyard,Rochette_etal,Soumyo,Louetal,Burgeretal}. Despite its popularity, an accurate and detailed analysis of the method is lacking in the literature. Recently, Soumyo {\it et al.} have analyzed the LW method in 1D and 2D for the linear convection-diffusion problem using GSA and they have identified the stability limits and acceptable range of simulation parameters for the method \cite{Soumyo}. They have also validated their analyses by solving 2D Navier-Stokes equation for the Taylor-Green vortex problem. Although the authors have performed detailed analyses with respect to the characterization of the method for explicit CD$_2$ scheme, important questions remain pertaining to its variants and general characteristics for multidimensional problems. 

It is known that for governing equations of the convection-diffusion type, two variants of the LW scheme can be derived based on the treatment of convection or convection-diffusion terms with the former variant analyzed in \cite{Soumyo}. However, these have not been analyzed in the literature with a view to compare/suggest the differences between the two. This is addressed in detail in the present manuscript using GSA and comparing/contrasting the relevant metrics. Also, it is important to note that the method has not been correctly analyzed with respect to error dynamics in the multidimensional space. This is also addressed in the present research where it is shown that the LW method includes an additional cross-derivative term which can impact its stability/accuracy properties. Finally, optimal parameters for the method are determined for accurate solution of the 2D Navier-Stokes equations and they are corroborated by computing the flow inside a square lid driven cavity at supercritical Reynolds numbers of 10,000. The computations with the present scheme are compared with available benchmark solutions \cite{NCCD1,NCCD2} which to the author's knowledge has never been done before.

The paper is organized as follows. In section \ref{gsa_sec}, the basic principles and rationale for GSA is presented. This is followed by section \ref{lw_1dce_sec} describing the Lax-Wendroff method for the 1D linear convection equation. In section \ref{lw_1dcde_sec}, the Lax-Wendroff method is derived for the 1D linear convection-diffusion equation and two of its variants are analyzed using GSA. Multidimensional analysis of the method is presented in section \ref{lw_2dcde_sec} for the 2D convection-diffusion equation and the optimal parameters are identified. The GSA results are corroborated in section \ref{NSEres_sec} by solving the 2D Navier-Stokes equations for the flow inside a square lid driven cavity and comparing the results with benchmark data. The paper ends with summary and conclusions in section \ref{sumcon_sec}.

\section{Quantifying Methods by Global Spectral Analysis (GSA)}
\label{gsa_sec}
In GSA, the unknown is represented by a hybrid-spectral form as,

\begin{equation}
u(x,t) = \int \hat{U}(k,t)e^{ikx}dk
\label{eqn6}
\end{equation}

Here $\hat{U}$ is the time-dependent Fourier amplitude and $k$ is the independent variable. The exact spatial derivative is obtained from Eq. \eqref{eqn6} as,
${\frac{\partial u}{\partial x} \biggl|}_{_{_{exact}}} = \int ik \hat{U} e^{ikx} dk$. 
One can write an equivalent numerical spatial derivative as, 
${\frac{\partial u}{\partial x}\biggl|}_{_{_{num}}} = \int ik_{eq} \hat{U} e^{ikx} dk$. 
The Fourier-Laplace amplitude is multiplied by $ik_{eq}$ for the numerical derivative (instead of $ik$ for the exact derivative). This notation for the numerical derivative was originally made popular by Vichnevetsky and Bowles \cite{HACM340} and subsequently in \cite{HACM153,B18,A1}, among many other references.

There are two main reasons for using $k_{eq}$ in the spectral representation: first, to provide the yardstick in comparing different discretization methods used in any computation by evaluating an equivalent $k_{eq}$ for the method. Expressions for this quantity have been presented for finite volume and Galerkin finite element methods for the solution of Eq. \eqref{eqn1} in \cite{HACM}. Ideally the ratio $k_{eq}/k$, should be equal to one and it is a measure of the resolution of the discretization scheme. One notes that this ratio is presented in the literature as a function of $kh$, with $h$ as the uniform grid spacing. The second use of $k_{eq}$ is due to its erroneous use for numerical dispersion relation. If one ignores error due to temporal discretization, then the form of the spatial derivative given in terms of $k_{eq}$ for Eq. \eqref{eqn1} has been used incorrectly by few researchers to write the numerical dispersion relation as,

\begin{equation}
\omega_{num} = k_{eq} c
\label{eqn7}
\end{equation}

This has also been noted as correct for the semi-discrete numerical stability analysis \cite{HACM39,HACM120,HACM360}. It may appear as  correct to treat $c$ as a constant in Eq. \eqref{eqn7}. However, a correct quantitative analysis is impossible by the semi-discrete approach using the numerical dispersion relation given by Eq. \eqref{eqn7}. It is noted that one can compute numerical group velocity for this approach following the basic definition as \cite{HACM,HACM329,LeVeque,Strikwerda},

\begin{equation}
v_{g,num} = \frac{d\omega_{num}}{dk} = c \frac{dk_{eq}}{dk}
\label{eqn8}
\end{equation}

However, the obtained group velocity is incorrect, as it is independent of time discretization scheme for Eq. \eqref{eqn1}. Propagation of wave-packet studied in \cite{HACM} show the group velocity is a strong function of both space and time discretizations. The authors in \cite{HACM312} have used this numerical dispersion relation in Eq. \eqref{eqn8} to derive a DRP scheme by considering spatial discretization alone. A four time-level method \cite{HACM312} was used to propose their DRP scheme, failing to note that such methods will invoke two spurious numerical modes, a topic explained in \cite{HACM,B93,B98}.

Instead of using Eq. \eqref{eqn7}, authors in \cite{Haltiner_Williams,HACM256} have shown that when one solves Eq. \eqref{eqn1}, the phase speed can no longer be a constant. This may appear paradoxical, but it is clearly explained in \cite{HACM256,HACM259,B93} that the space-time discretization method fixes the phase shift per time step, which in turn determines the numerical phase speed ($c_{num}$), that is different from the physical phase speed, $c$. Then $c_{num}$ becomes function of wavenumber/circular frequency, giving rise to phase and dispersion error. This simple, yet subtle cause for $c_{num} \neq c$, is one of the central results of GSA in the expression for the numerical dispersion relation given in Eq. \eqref{eqn5}. The correct group velocity accounting for the space-time discretization is therefore given by,

\begin{equation}
v_{g,num} = c_{num} + k \frac{dc_{num}}{dk}
\label{eqn9}
\end{equation}

We note that $k$ is the independent variable, and the spatial and temporal discretization schemes fix the numerical dispersion relation, which in turn fixes other numerical parameters and error dynamics \cite{HACM, B94}. The fact that $c$ changes to $c_{num}$, applies equally to other coefficients of many other transport and diffusion equations \cite{B94,B118,Focusing_CISM}.

\subsection{Rationale for the GSA}

To justify the use of GSA, an explanation is provided with the help of an example to demonstrate the utility of Eq. \eqref{eqn5} for the 1D convection equation solved by midpoint leap-frog scheme \cite{LFR22} on a uniform grid of spacing $h$. Discretizing Eq. \eqref{eqn1}, one obtains the difference equation for a node at ($x_j, t^n$) as,

\begin{equation}
\frac{u^{n+1}_j - u^{n-1}_j}{2\Delta t} + c\left(\frac{u^{n}_{j+1}-u^{n}_{j-1}}{2h}\right) = 0
\label{disc_lf_cd2}
\end{equation}

\noindent where $\Delta t$ is the time-step used. We note that the numerical phase shift derived using Eq. \eqref{eqn6} is different from that given by the $\Pi$-form analysis \cite{HACM286}, due to the fact that the numerical dispersion relations of these two approaches are different.

For the discrete Eq. \eqref{disc_lf_cd2}, the unknown is next represented by Eq. \eqref{eqn6}. In the $\Pi$-form analysis, the numerical phase speed is computed from the dispersion relation as $c_{{num}_\Pi}=\frac{k_{eq}}{k}c$ where $\omega_\Pi = kc_{{num}_\Pi}$. For the second order central difference scheme, $k_{eq}=\frac{\sin(kh)}{h}$. Thus, the numerical phase speed from $\Pi$-form for the leap-frog scheme is given by

\begin{equation}
\frac{c_{{num}_\Pi}}{c}=\frac{\sin(kh)}{kh}
\end{equation}

The $\Pi$-form analysis in \cite{HACM340} reports the above expression (cf. Eq. (2.13)). There are two distinctive features of this result. First, the time-integration is by a three-time level method, and one gets
 two distinct numerical phase speeds (not given here). Secondly, and most importantly, the assumption that the numerical phase speed is independent of time discretization is used in this $\Pi$-form analysis. A correct approach based on the $\Gamma$-form approach is given next.

The circular frequency and phase speed is obtained as per the governing differential equation and its discretization in writing the numerical dispersion relation by Eq. \eqref{eqn5}. Based on the difference equation, one obtains the numerical amplification factor following the representation in Eq. \eqref{eqn6}. Such a numerical amplification factor fixes phase shifts per time step, and provides the numerical phase speeds. For the $\Gamma$-form analysis, the unknown is represented by Eq. \eqref{eqn6}. One can represent the initial condition for Eq. \eqref{eqn1} as,

\begin{equation}
u(x_j,t = 0) = u_j^0 = \int \hat{U}_0 (k)\; e^{ikx_j} dk
\label{ini_cond}
\end{equation}

\noindent with the subscript and superscript denoting the spatial and temporal indices, respectively. The solution at any time, $t = n \Delta t$, is written using the numerical amplification factor as

\begin{equation}
u_j^n = \int \hat{U}_0(k)\; |G_{num,j}|^n\; e^{i(kx_j - n\phi_j)} dk
\label{anytime}
\end{equation}

\noindent where $G_{num,j} = \left(\frac{\hat{U}(k,t^n+\Delta t)}{\hat{U}(k,t^n)}\right)$ is the complex amplification factor, i.e. $G_{num,j} = G_{num,rj} + iG_{num,ij}$, such that
$|G_{num,j}| = (G_{num,rj}^2 + G_{num,ij}^2)^{1/2}$. The phase shift per time step is calculated as $\tan \phi_j = -{G_{num,ij}}/{G_{num,rj}}$. From this the numerical phase speed ($c_{{num}_\Gamma}$) is obtained as,

\begin{equation}
c_{{num}_\Gamma}=\frac{\phi_j}{k\Delta t}
\label{cnrel}
\end{equation}

Here, $c_{{num}_\Gamma}$ depends on $k$, i.e. the numerical solution is dispersive, as opposed to the non-dispersive nature of physical solution. Thus, both the $\Pi$- and $\Gamma$-form analyses show the numerical phase speed to depend on $k$, with the main difference that the latter uses the temporal discretization information. In contrast, the $\Pi$-form analysis (like the semi-discrete analysis) ignores any information originating from the temporal discretization.

Substituting the variable $u$ given by Eq. \eqref{eqn6} in Eq. \eqref{disc_lf_cd2}, one obtains the following,

\begin{equation}
\hat{U}^{n+1}_j-\hat{U}^{n-1}_j + \frac{c\Delta t}{h}\left(e^{ikh} - e^{-ikh}\right) \hat{U}^n_j = 0
\end{equation}

\noindent where the variables with the hat denote spectral amplitudes. With the definition of numerical amplification factor $G_{num,j}=\left(\frac{\hat{U}^{n+1}_j}{\hat{U}^n_j}\right) = \left(\frac{\hat{U}^{n}_j}{\hat{U}^{n-1}_j}\right)$, a quadratic for $G_{num,j}$ is obtained with the two roots obtained as,

 \begin{equation}
{G_{num,j}}_{1,2} = -iN_c\sin(kh) \pm \sqrt{1-{N_c}^2 \sin^2(kh)}
\end{equation}

\noindent where $N_c = \frac{c \Delta t}{h}$ is the Courant-Friedrich-Lewy (CFL) number. From the above, $c_{{num}_\Gamma}$ is calculated from Eq. \eqref{cnrel} as

 \begin{equation}
\frac{c_{{num}_{\Gamma_{1,2}}}}{c} =\frac{\phi_j}{(kh) N_c}= \left(\frac{1}{(kh)N_c}\right) \tan^{-1}\left({\frac{N_c \sin(kh)}{\pm \sqrt{1- {N_c}^2 \sin^2(kh)}}}\right)
\label{cn_gsa}
\end{equation}

Vichnevetsky and Bowles \cite{HACM340} also used the $\Gamma$-form analysis to determine the numerical phase speed. The unknown is represented by the Fourier-Laplace transform and substituting in Eq. \eqref{disc_lf_cd2}, one obtains the spectral plane representation for the leap-frog scheme as,

$$e^{-i\omega_0 \Delta t} - e^{i\omega_0 \Delta t} + N_c\left( e^{ikh} - e^{-ikh}\right) = 0$$
which upon simplification yields,
$\sin(\omega_0 \Delta t) = N_c\sin(kh)$.

This provides the circular frequency as,
\begin{equation}
\omega_0 =\frac{1}{\Delta t} \sin^{-1}(N_c\sin(kh))
\label{WDR}
\end{equation}

With the expression for $\omega_0$ in terms of $k$, numerical phase speed is computed as, $c_{{num}_{VB}} = \frac{\omega_0}{k}$, which upon simplification yields

\begin{equation}
\frac{c_{{num}_{VB}}}{c}=\frac{1}{(kh)N_c}\sin^{-1}(N_c\sin(kh))
\label{cVB}
\end{equation}

We note that Vichnevetsky and Bowles obtained the amplification factors given by Eq. (4.8c) in \cite{HACM340} and show the above expression for numerical phase speed in Table 4.3 \cite{HACM340}. At a first glance the numerical phase speed computed from the $\Gamma$-form analysis (Eq. \eqref{cn_gsa}) and the expression given by Vichnevetsky and Bowles \cite{HACM340} (Eq. \eqref{cVB}) may appear as different. But, these are equivalent following the trigonometric identity: $\sin^{-1}(x) = \tan^{-1}\left( \frac{x}{\sqrt{1-x^2}}\right)$.

Having described the physical and different form of numerical dispersion relations, and appearance of additional spurious numerical modes for three/ higher time level methods for solving Eq. \eqref{eqn1}, we discuss about the LW method, which is unambiguous and helps explain some of the fundamental principles of the MDE.

\section{The Lax-Wendroff Method for the Convection Equation}
\label{lw_1dce_sec}
In this method, the proponents retained second order accuracy for time integration in discretizing the governing equation. For the MDE, we have an example for which spurious modes are not invoked, while the time integration is by a higher order method, with the DRP property improved. The difference equation is obtained from the Taylor series expansion,

\begin{equation}
u_j^{n+1}= u_j^n + \Delta t \frac{\partial u}{\partial t} + \frac{(\Delta t)^2}{2} \frac{\partial^2 u}{\partial t^2}
\label{LWS1}
\end{equation}

In the LW method, the second time derivative is obtained from Eq. \eqref{eqn1} as,
$\frac{\partial^2 u}{\partial t^2} = c^2 \frac{\partial^2 u}{\partial x^2}$. When the first and second time derivatives are substituted in Eq. \eqref{LWS1}, one obtains the MDE given by,

\begin{equation}
u(t + \Delta t)= u(t) -c \Delta t \frac{\partial u}{\partial x} + \frac{c^2 (\Delta t)^2}{2} \frac{\partial^2 u}{\partial x^2}
\label{LWS2}
\end{equation}

One immediately notices the presence of the last term on the right hand side as a strictly diffusive term that was absent in the original governing equation. This shows that a pure convection equation (a hyperbolic PDE) is converted into CDE (a parabolic PDE). This added diffusive term may provide numerical stabilization, as often used in CFD. However by Lax's theorem, this is an issue of inconsistency, as a hyperbolic PDE is converted to a parabolic PDE \cite{WFAmes,HACM}. This has been noted for Dufort-Frankel method \cite{HACM65,HACM} also, where the heat equation (parabolic PDE) is converted into a wave propagation problem -- another example of inconsistency, if the numerical parameters are not chosen with care (see Fig. 6.4 in \cite{HACM} for details). In defense of the LW method, one should note that with refined time steps, the inconsistency due to added diffusion can be made progressively sub-dominant.

\section{The Lax-Wendroff Method for the 1D Convection-Diffusion Equation}
\label{lw_1dcde_sec}
In the previous section, the LW method is presented for the 1D convection equation and a key issue of inconsistency of the MDE is demonstrated. Here, we evaluate using GSA, the same method applied to a 1D convection-diffusion equation (CDE). The objective in doing this study is two-fold. First objective is to check if the MDE is consistent with the original PDE. The second objective is to briefly analyze and contrast two forms of application of the LW method- application to only convection term and both convection-diffusion terms, respectively. The former strategy is employed by a popular and well established research code- ABVP, developed at CERFACS, France, for problems of combustion \cite{Wang_etal,Brunet_etal,Schoenfeld_Rudgyard,Rochette_etal}. 
%

However, no analysis or quantification is provided in the literature to compare the above two strategies. This is established here by using GSA and comparing the numerical properties of the two strategies for representative simulation parameters. This is one of the contributions of the present work and has never been done before.     
   
We note that unlike for the convection equation, if the LW method is used in the 1D CDE, the problem of inconsistency is not present as shown next. However, the effects of additional diffusion terms will determine the accuracy of space-time discretization simultaneously. For the CDE, the governing canonical equation is given by,

\begin{equation}
\frac{\partial u}{\partial t} + c\frac{\partial u}{\partial x} = \alpha \frac{\partial^2u}{\partial x^2}
\label{eqn22}
\end{equation}

For the LW method substitution of Eq. \eqref{eqn22} in Eq. \eqref{LWS1} yields the following,

\begin{equation}
u(t + \Delta t)= u(t) -c \Delta t \frac{\partial u}{\partial x} + \alpha \Delta t \frac{\partial^2 u}{\partial x^2} + \frac{(\Delta t)^2}{2} \frac{\partial^2 u}{\partial t^2}
\label{eqn23}
\end{equation}

Furthermore, deriving $\frac{\partial^2 u}{\partial t^2}$ from Eq. \eqref{eqn22} one obtains

\begin{equation}
 \frac{\partial^2 u}{\partial t^2} = c^2 \frac{\partial^2 u}{\partial x^2} - 2 \alpha c \frac{\partial^3 u}{\partial x^3} + \alpha^2 \frac{\partial^4 u}{\partial x^4}
\label{eqn24}
\end{equation}

Substitution of Eq. \eqref{eqn24} in Eq. \eqref{eqn23}, one notices added diffusive terms arising from the pure convection and diffusion terms. Additionally there is a dispersive term due to the interaction between convection and diffusion terms. A higher order diffusion term  also appears from the basic diffusion term, which can act like the hyper-viscosity term added in pseudo-spectral method to provide numerical stabilization \cite{WSEAS}, when used with lower order Runge-Kutta schemes.

\subsection{The Lax-Wendroff method for explicit central difference scheme}

One can use central difference schemes for various derivatives in Eqs. \eqref{eqn22} and \eqref{eqn24}, and substituting these in Eq. \eqref{eqn23}, one gets the corresponding difference equation. Using the following notations: $\frac{\partial^2 u}{\partial x^2} = h^2 D^2 u$, $\frac{\partial^3 u}{\partial x^3} = h^3 D^3u$ and $\frac{\partial^4 u}{\partial x^4} = h^4 D^4u$, and the non-dimensional parameters: $N_c=\frac{c\Delta t}{h}$ and $Pe = \frac{\alpha \Delta t}{h^2}$, as the CFL and the Peclet numbers, the difference equation becomes,

\begin{equation}
 u_j^{n+1} = u_j^n - \frac{N_c}{2} (u^n_{j+1} - u^n_{j-1}) +(Pe +\frac{N_c^2}{2}) (u^n_{j+1} -2u_j^n + u^n_{j-1})- Pe N_c D^3 u_j^n + \frac{Pe^2}{2} D^4u_j^n
\label{eqn25}
\end{equation}

Various differences appearing in Eq. \eqref{eqn25} can be evaluated using full and half-node locations of grid points. For example,  half-node locations provide the following, $D^2u_j = u_{j-1} -2u_j + u_{j+1}$ and $D^4u_j = 16(u_{j+1} -4u_{j+1/2} +6u_j -4u_{j-1/2}+u_{j-1})$. For $D^3u_j$, a combination of full and half-node representations provide, $D^3u_j= (u_{j+2}-2u_{j+1}+2u_{j-1} -u_{j-2})/2$. A formal Taylor series expansion would reveal the consistency of these difference expressions. Note that in some applications, the LW method is used for convection terms only \cite{Wang_etal,Brunet_etal,Schoenfeld_Rudgyard,Rochette_etal}. In such cases, one would switch off the $D^3u_j$ and $D^4u_j$ terms. In the following, the numerical properties of methods to solve the CDE, with and without $D^3u_j$ and $D^4u_j$ terms, are compared following the analysis given in \cite{B94}.

For the 1D CDE, $c$ and $\alpha$ are constant real numbers and for a generic non-periodic problem one can represent $u$ by Eq. \eqref{eqn6} so that the CDE becomes,

\begin{equation}
\frac{d\hat{U}}{dt} + ick\hat{U} = -\alpha k^{2}\hat{U}.
\label{eqn26}
\end{equation}

\noindent which can be solved analytically for a general initial condition in Eq. \eqref{ini_cond} to yield,
\begin{equation}
 \hat{U}(k,t) = \hat{U}_0(k) \: e^{-\alpha k^{2}t} e^{-ikct}.
 \label{eqn27}
\end{equation}

The physical dispersion relation of the CDE is obtained using Eq. \eqref{eqn3} as,

\begin{equation}
 \omega_0 = c \: k - i \:\alpha \: k^{2}.
 \label{eqn28}
\end{equation}

The physical phase speeds is obtained from,
\begin{equation}
  c_{phys} = \frac{\omega_0}{k} =  c - i \: \alpha k,
  \label{eqn29}
\end{equation}

The phase speed is taken as the real part of the above, as reported in the literature. The physical group velocity is the energy propagation velocity of a wave-packet and is given following the definition in \cite{HACM,HACM329} as,

\begin{equation}
  v_{g,phys} = \frac{d\omega_0}{dk} = c - 2 \: i \: \alpha k.
  \label{eqn30}
\end{equation}

The implication of this complex group velocity is given in \cite{B94} and for real coefficient of diffusion, one obtains the real part of the group velocity given by, $c$, and the same is used here for the CDE.

The analytical solution of the CDE can be interpreted by the physical amplification factor, i.e. comparing the solution amplitude at two distinct instants separated by $\Delta t$ so that,

\begin{equation}
  G_{phys} = \frac{\hat{U}(k,t+\Delta t)}{\hat{U}(k,t)} =
  e^{-\alpha \: k^{2}\Delta t} e^{-i \: k \: c \: \Delta t} = e^{-i \: \omega_0\: \Delta t }
  \label{eqn31}
\end{equation}

\noindent which can be expressed in terms of the numerical parameters, $N_c$ and $Pe$ as,

\begin{equation}
  G_{phys} = e^{-Pe \: (kh)^{2}} \: e^{-i \: N_c \: (kh)}.
  \label{eqn32}
\end{equation}

It has been shown in \cite{B94,B118,B119} that contrary to the popular assumption of von-Neumann analysis, the numerical convection speed, $c_{num}$, and the numerical diffusion, $\alpha_{num}$, are dependent on $k$, $h$, $N_c$ and $Pe$. All numerical methods have their numerical amplification factor, $G_{num}$, and numerical dispersion relation which dictate the evolution of the solution in time. It is essential that for any numerical scheme, $G_{num}$ should be as close as possible to $G_{phys}$.

The numerical discretization of the continuous problem would result in a numerical dispersion relation equivalent to Eq.~(\ref{eqn28}) as derived in~\cite{B94}, and expressed as,

\begin{equation}
  \omega_{num} = c_{num} \: k - i \: \alpha_{num} \: k^{2}.
  \label{eqn33}
\end{equation}

\noindent where $\omega_{num}$ is complex and differs from the physical dispersion expression, since $c_{num}$ and $\alpha_{num}$ vary with $kh$, $N_c$ and $Pe$. Exploiting the similarity to the exact solution, the numerical amplification factor, $G_{num}$ is obtained as a function of $\alpha_{num}$ and $c_{num}$ as,

\begin{equation}
  G_{num} = e^{-\alpha_{num} \: k^{2}\Delta t} \: e^{-i \: k \:c_{num} \Delta t} = e^{-i \: \omega_{num} \Delta t}.
  \label{eqn34}
\end{equation}

From Eq. \eqref{eqn25}, for the LW method based on explicit central differences for the 1D CDE, one obtains the complex $G_{num}$ as

\begin{equation}
\begin{split}
 G_{num} = \,& 1-\frac{N_c}{2}\left( e^{ikh}-e^{-ikh}\right) +\left(\frac{N_c^2}{2}+Pe\right)\left(e^{ikh}-2+e^{-ikh}\right) \\[1.5ex]
 &-\frac{Pe\,N_c}{2}\left( e^{2ikh}-2e^{ikh}+2e^{-ikh}-e^{-2ikh}\right)\\[1.5ex]
  &+8 Pe^2\left( e^{2ikh}-4e^{ikh/2}+6-4e^{-ikh/2}+e^{-2ikh}\right)
\end{split}
\label{gn_cd2}
\end{equation}

Using similar previous relations, one notes that:

\begin{equation}
  c_{num} = \Re\left(\frac{\omega_{num}}{k}\right),   \label{eqn35}
\end{equation}
and,
\begin{equation}
   v_{g,num} = \Re\left(\frac{d\omega_{num}}{dk}\right).
   \label{eqn36}
\end{equation}

One can evaluate $c_{num}$ from the numerical phase shift per time step, as the ratio between the imaginary and real part of $G_{num}$ given by,

\begin{equation}
  tan(\beta) = -\frac{\Im(G_{num})} {\Re(G_{num})}
  =  tan(c_{num} k \Delta t).
\label{eqn37}
\end{equation}
\noindent where the real and imaginary parts of $G_{num}$ are represented in the above as $\Re(G_{num})$ and $\Im(G_{num})$, respectively. Also, note that for the discussion of convection equation we have termed this phase shift per time step as $\beta$.

This allows one to express the non-dimensional effective numerical phase speed as,

\begin{equation}
  \frac{c_{num}}{c_{phys}} = \frac{\beta}{kc\Delta t} =
-\frac{1}{(kh)N_c} \: tan^{-1}\left[\frac{\Im(G_{num})} {\Re(G_{num})}\right].
 \label{eqn38}
\end{equation}

Similarly, the numerical group velocity is obtained as,

\begin{equation}
v_{g,num} = \Re\left( \frac{\partial \omega_{num}}{\partial k}\right) = \frac{1}{\Delta t}{\partial \beta \over \partial k},
\label{eqn39}
\end{equation}

\noindent which can be written in normalized form as,

\begin{equation}
 \frac {v_{g,num}} {v_{g,phys}} = \frac{1}{N_c}\frac{d\beta}{d(kh)}.
 \label{eqn40}
\end{equation}

Following Eq.~\eqref{eqn32}, one directly notes that the modulus of numerical amplification factor is solely dependent on $\alpha_{num}$ by analogy as,

\begin{equation}
   |G_{num}| = e^{-\alpha_{num} k^{2}\Delta t},
   \label{eqn41}
\end{equation}

Alternately in terms of $Pe$, one can write this modulus also as,

\begin{equation}
    ln|G_{num}| = -\frac{\alpha_{num}} {\alpha} (kh)^{2} \: Pe.
    \label{eqn42}
\end{equation}

Equation~\eqref{eqn42} can also be used to evaluate the numerical diffusion coefficient in non-dimensional form as,

\begin{equation}
    \frac {\alpha_{num}}{\alpha} = -\frac {ln|G_{num}|}{(kh)^{2} Pe}.
    \label{eqn43}
\end{equation}

This clearly establishes that a numerical scheme for the CDE changes the diffusion process differently for different length scales ($kh$), depending on the numerical scheme. For the LW method this corresponds to choosing the spatial discretization only, and that is a special feature of the LW method. From Eq. \eqref{eqn25} one notes that the retention of higher order term for the time integration will also involve the terms $D^3u_j^n$ and $D^4u_j^n$, when the diffusion term is expressed with the second order time discretization.

Using Eqs. \eqref{gn_cd2}, \eqref{eqn43}, \eqref{eqn38} and \eqref{eqn40}, the properties $\frac{|G_{num}|}{|G_{phys}|}$, $\frac{\alpha_{num}}{\alpha}$, $\frac{c_{num}}{c_{phys}}$ and $\frac{v_{g,{num}}}{v_{g,phys}}$ are obtained for any ($N_c$, $Pe$) combination. These property charts are then used to gauge the accuracy of the developed scheme.

    \begin{figure}
        \centering
         \includegraphics[scale=0.55]{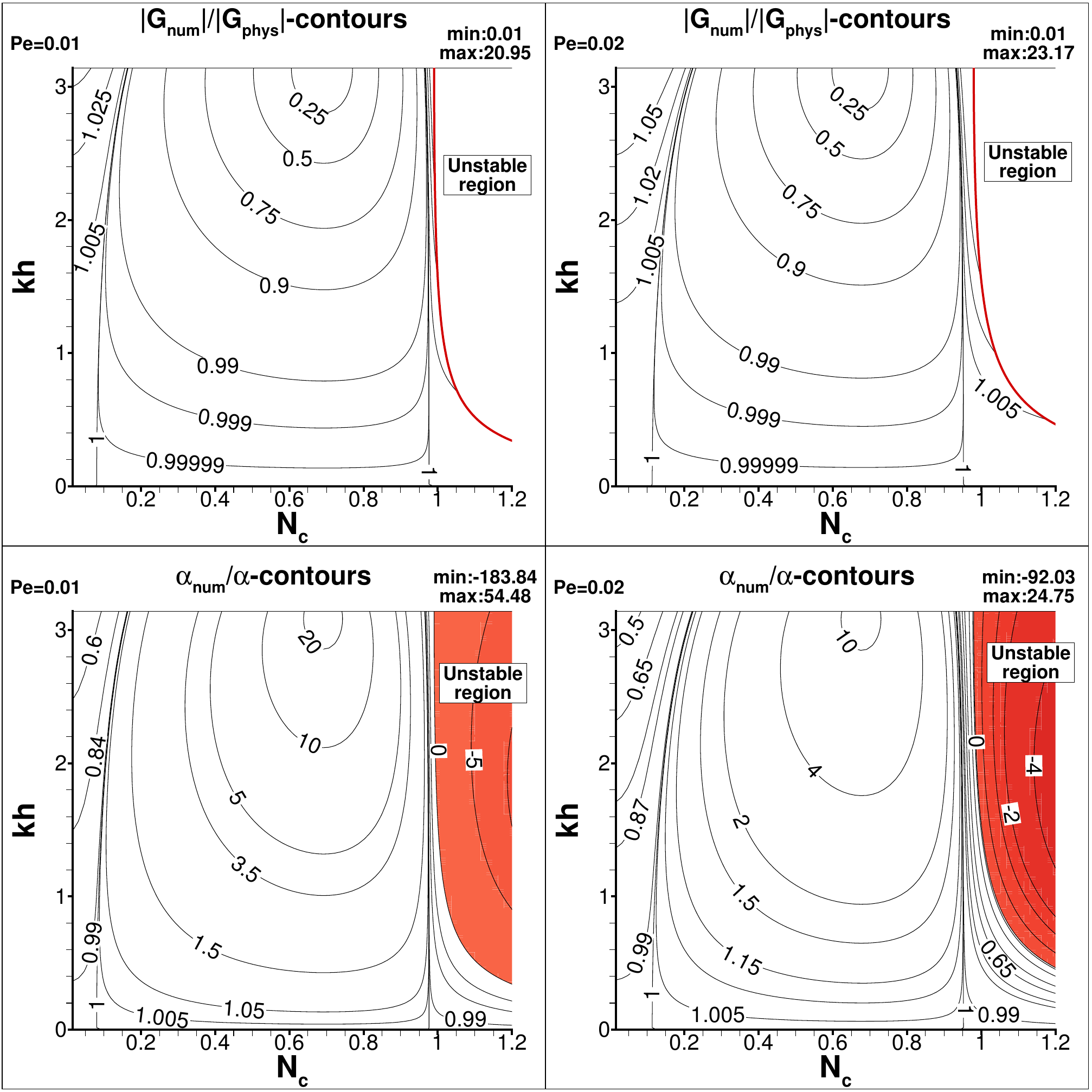}
        \caption{The ratio of numerical and physical amplification factors $\left(\frac{|G_{num}|}{|G_{phys}|}\right)$ and the ratio of numerical and the physical diffusion coefficients $\left(\frac{\alpha_{num}}{\alpha}\right)$ for the Lax-Wendroff method based on explicit CD scheme applied on the convection term only of the CDE plotted in the $(N_c,kh)$-plane for the representative Peclet numbers $Pe=0.01$ and $0.02$. Regions of numerical instability are as marked in the panels.}
        \label{fig1}
    \end{figure}

    \begin{figure}
        \centering
         \includegraphics[scale=0.55]{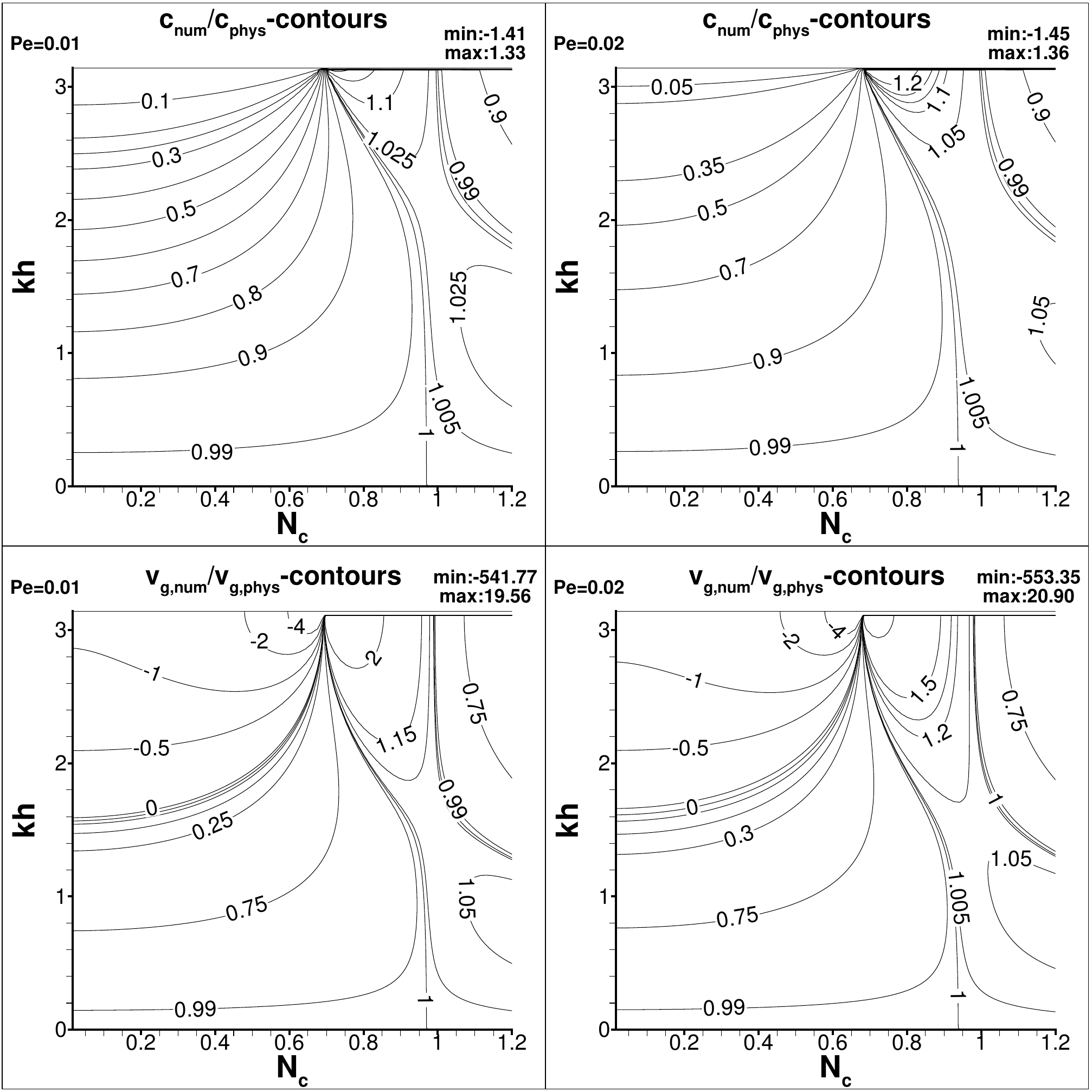}
        \caption{The ratio of numerical and physical phase speeds $\frac{c_{num}}{c_{phys}}$ and the ratio of numerical and the physical group velocity $\frac{v_{g,num}}{v_{g,phys}}$ for the Lax-Wendroff method based on explicit CD scheme applied on the convection term only of the CDE plotted in the $(N_c,kh)$-plane for the representative Peclet numbers $Pe=0.01$ and $0.02$.}
        \label{fig2}
    \end{figure}

In Figs. \ref{fig1} and \ref{fig2}, property charts are shown in the $(N_c,kh)$-plane for the Peclet numbers, $Pe=0.01$ and $0.02$, for the LW method based on explicit central differences as applied on the convection term of the governing CDE, i.e. $D^3$ and $D^4$ terms in Eq. \eqref{eqn25} are not included. From Fig. \ref{fig1}, one notes that at low $N_c$ values, the ratio $\frac{|G_{num}|}{|G_{phys}|}$ is greater than unity for all $k$, implying that the numerical solution has lower numerical diffusion. This is corroborated from the plot of $\frac{\alpha_{num}}{\alpha}$ contours where the corresponding values are lower than one. Numerical diffusion is also noted to decrease progressively with increasing $k$. However adjacent to this region, another region exists with higher numerical diffusion compared to the physical value at all wavenumbers with the diffusion strength increasing with $k$. Beyond $N_c=1$, $\alpha_{num}$ is negative implying presence of anti-diffusion for some range of resolved $k$ that will lead to catastrophic numerical instability as shown in the marked regions in red in the bottom frames of Fig. \ref{fig1}.

The numerical phase speed ($c_{num}$) and group velocity ($v_{g,num}$) contours plotted in Fig. \ref{fig2}, show poorer performance of the LW scheme in preserving a signal's physical dispersion and propagation characteristics as compared to many compact schemes used with higher order Runge-Kutta methods \cite{HACM}. This performance degradation is attributed to the poor resolution of explicit CD scheme in calculating the spatial derivatives. However, some specific ranges of $kh$ and $N_c$ can be found, where this method can be used for very low error tolerance. This is one of the reasons for the reported analysis here, which helps in locating specific $N_c$ and $kh$ ranges for a fixed $Pe$.

    \begin{figure}
        \centering
         \includegraphics[scale=0.55]{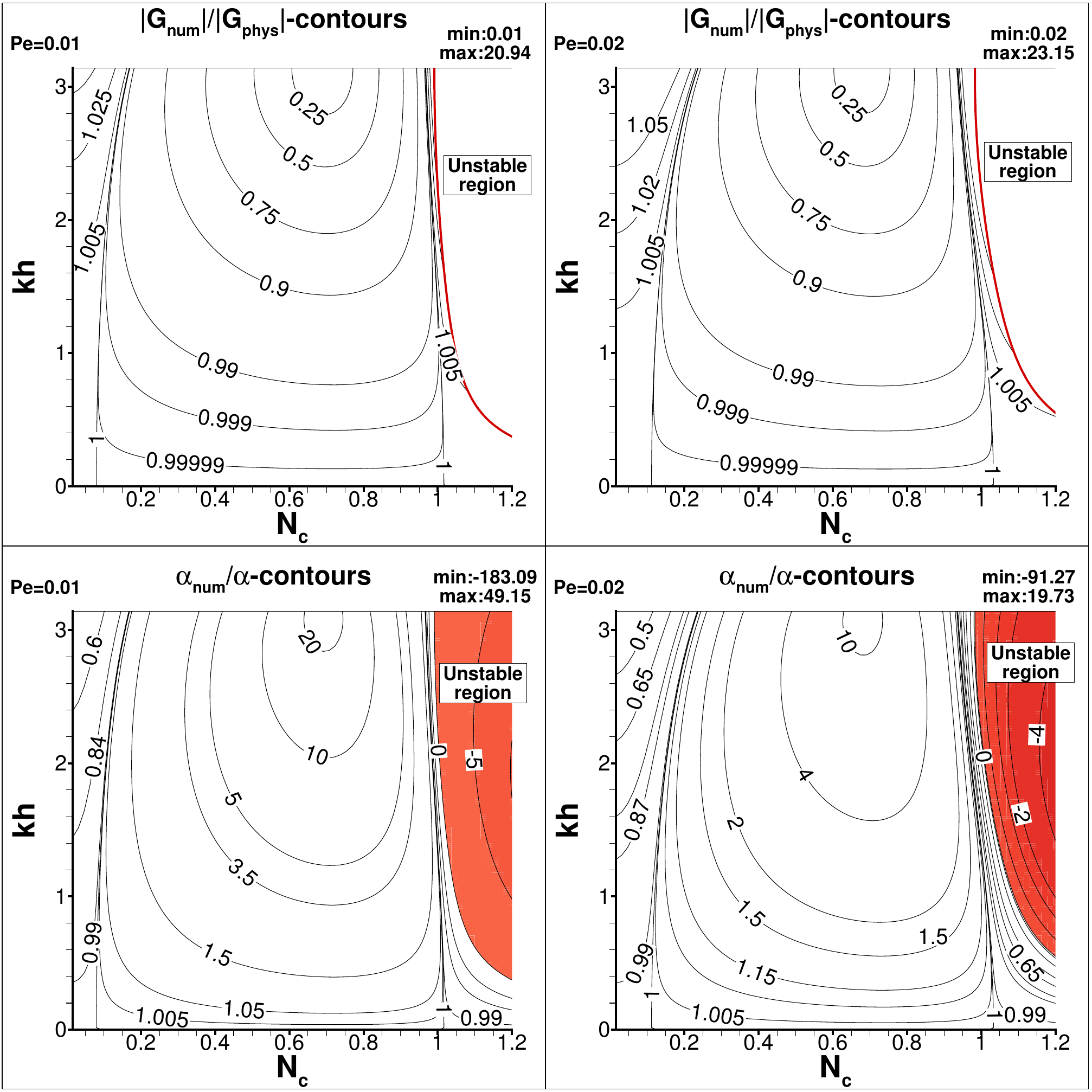}
        \caption{The ratio of numerical and physical amplification factors $\left(\frac{|G_{num}|}{|G_{phys}|}\right)$ and the ratio of numerical and the physical diffusion coefficients $\left(\frac{\alpha_{num}}{\alpha}\right)$ for the Lax-Wendroff method based on explicit CD stencils for D3 and D4 terms included in CDE, plotted in the $(N_c,kh)$-plane for the representative Peclet numbers $Pe=0.01$ and $0.02$. Regions of numerical instability are as marked in the panels.}
        \label{fig3}
    \end{figure}

    \begin{figure}
        \centering
         \includegraphics[scale=0.55]{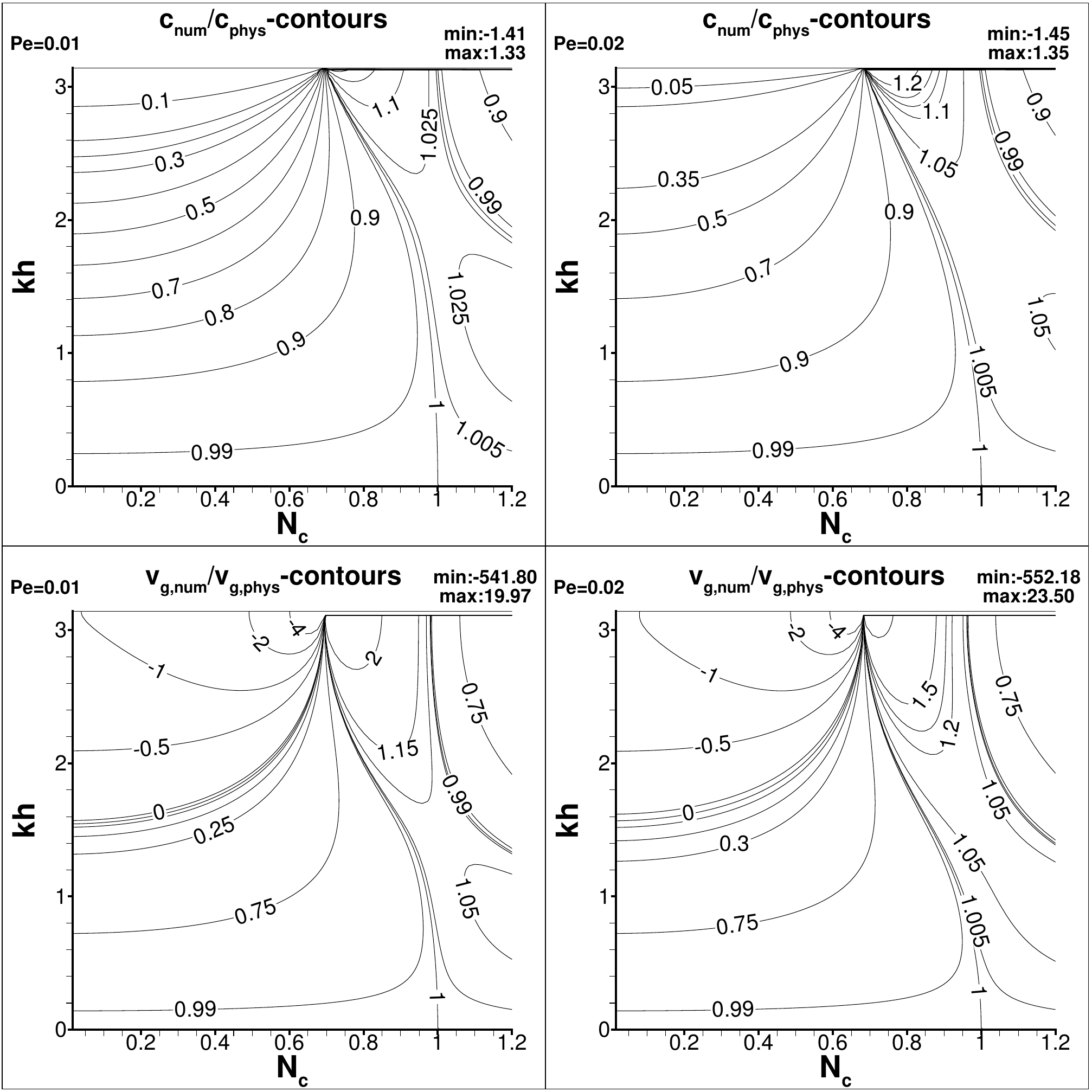}
        \caption{The ratio of numerical and physical phase speeds $\frac{c_{num}}{c_{phys}}$ and the ratio of numerical and the physical group velocity $\frac{v_{g,num}}{v_{g,phys}}$ for the Lax-Wendroff method based on explicit CD stencils for D3 and D4 terms included in CDE plotted in the $(N_c,kh)$-plane for the representative Peclet numbers $Pe=0.01$ and $0.02$.}
        \label{fig4}
    \end{figure}

The relative advantages of using the LW method including the $D^3u$ and $D^4u$ terms, based on explicit central differences in the CDE are noted in Figs. \ref{fig3} and \ref{fig4}, with property charts plotted in the $(N_c,kh)$-plane for the same two $Pe$ values used in Figs. \ref{fig1} and \ref{fig2}. As noted earlier for the method without $D^3u$ and $D^4u$ terms, one notes similar behavior for $\alpha_{num}$, i.e. lower $\alpha_{num}$ occurring at lower $N_c$ values, whereas higher $\alpha_{num}$ occurs for an adjacent range of $N_c$, as compared to the physical value of $\alpha$. Comparing Figs. \ref{fig1} and \ref{fig3}, one notes the LW method with $D^3u$ and $D^4u$ terms to have an increased stable region, as compared to the LW method applied only for the convection term. The full LW scheme also has better numerical diffusion at all $kh$. This is a consequence of retaining the $D^4u$ term that moderates the added/reduced numerical diffusion across the $N_c$ range. The role of positive fourth diffusion term in Eq. \eqref{eqn25} can be understood by noting that $\partial^4 u \over \partial x^4$ term varies as $k^4$ in the spectral plane, thereby reducing the added numerical diffusion by the introduced second derivative term in the Taylor series expansion of the convection term.

Comparing the numerical phase speed and group velocity contours between the convection alone and full LW method based on CD schemes from Figs. \ref{fig2} and \ref{fig4}, one notes the latter to have properties extended slightly in the $N_c$ direction due to increased stability. Apart from this, only marginal differences are noted in terms of accuracy between the two strategies for dispersion errors.

We note that both strategies with CD scheme are less suitable for high accuracy solutions of the CDE due to present errors at all $kh$ noted for $\alpha_{num}$, $c_{num}$ and $v_{g,num}$. This implies that these strategies are strictly less suitable for DNS, due to noted one-to-one correspondence between the 2D linear CDE and 2D Navier-Stokes equations demonstrated in \cite{B118}. However, it is feasible to quantify the parameters for the LW method on the CDE and look for some combinations of $N_c$ and $kh$ ranges, which is shown next for the various differences appearing in the LW method used for the 2D CDE.

\section{Quantifying LW method using 2D CDE for LES}
\label{lw_2dcde_sec}
Having established that the convection-only LW discretization of the 1D CDE is comparable in accuracy to the LW method retaining $D^3$, $D^4$ terms, in this section, the properties of the same scheme are further analyzed for the 2D CDE. Previous research work \cite{B118} has demonstrated a one-to-one correspondence of the linear 2D CDE with the 2D Navier-Stokes equation. Hence, its role to analyze schemes for the Navier-Stokes equation cannot be over-emphasized, and is adopted here for the same purpose. 

The essentials of the GSA of 2D CDE required to analyze the 2D LW scheme are given below. Interested readers can consult \cite{B118} for details and discussion regarding the expressions obtained from GSA.  

The 2D CDE is given as,

\begin{equation}
\frac{\partial u}{\partial t} + c_x\frac{\partial u}{\partial x} + c_y\frac{\partial u}{\partial y} = \alpha \left(\frac{\partial^2u}{\partial x^2} + \frac{\partial^2u}{\partial y^2}\right)
\label{eqn2DCDE}
\end{equation}

For the 2D CDE, the LW scheme applied to the convection terms can be obtained as

\begin{equation}
u(t + \Delta t)= u(t) - \Delta t \left(c_x \frac{\partial u}{\partial x}+ c_y \frac{\partial u}{\partial y}\right)+ \alpha \Delta t \left(\frac{\partial^2 u}{\partial x^2}+\frac{\partial^2 u}{\partial y^2}\right) + \frac{(\Delta t)^2}{2} \left(c_x^2\frac{\partial^2 u}{\partial x^2} + 2 c_x c_y \frac{\partial^2 u}{\partial x\partial y} + c_y^2\frac{\partial^2 u}{\partial y^2}\right)
\label{2DLW}
\end{equation}

To perform GSA of Eq. \eqref {eqn2DCDE}, the unknown is expressed in the hybrid-spectral plane as,
\begin{equation}
u(x,y,t)= \iint \hat{U}(k_x,k_y,t)e^{ i(k_x x + k_y y)}dk_x dk_y
\end{equation}
\noindent where $\hat{U}$ is the Fourier-Laplace amplitude and $k_x$, $k_y$ are the wavenumber components in the $x$- and $y$-directions, respectively. Substituting the expression for $u$ in the governing equation one obtains, 
\begin{equation}
\hat{U}(k_x,k_y,t) = \hat{U}_0(k_x,k_y) \; e^{-\alpha(k_x^2 + k_y^2)t}e^{- i(k_x c_x + k_y c_y)t}
\end{equation}

\noindent where $u(x,y,0) = \iint \hat{U}_0(k_x,k_y)e^{ i (k_x x + k_y y)} dk_x dk_y$ is the initial solution.

The physical dispersion relation, an important property that must be obeyed by numerical schemes to minimize phase and dispersion errors, is obtained by expressing $u$ in the full spectral space as
\begin{equation}
\omega_0 = c_x k_x + c_y k_y -  i \alpha(k_x^2 + k_y^2)
\label{omega}
\end{equation}

From this relation the physical phase speed is obtained as,
\begin{equation}
c_{phys} = \frac{\omega_0}{\sqrt{k_x^2 + k_y^2}}=  \frac{ c_x k_x + c_y k_y}{\sqrt{k_x^2 + k_y^2}}
\label{comp_phase}
\end{equation}

\noindent and the physical group velocity components are obtained as,
\begin{equation}
v_{gx,phys} = \frac{\partial \omega_0}{\partial k_x} = c_x
\end{equation}
\begin{equation}
v_{gy,phys} = \frac{\partial \omega_0}{\partial k_y} = c_y
\end{equation}

The physical amplification factor for the CDE is 
\begin{eqnarray}
G_{phys} = e^{-[Pe_x(k_xh_x)^2 + Pe_y(k_yh_y)^2]}e^{- i[N_{cx}k_xh_x + N_{cy}k_yh_y]}
\label{Gphy} 
\end{eqnarray}
\noindent where $\Delta t$ is the discrete time-step and $h_x$ and $h_y$ are the grid spacings in $x$- and $y$-directions, respectively. The CFL and Peclet numbers in 2D are:  
${N}_{cx} = \frac {c_x \Delta t}{h_x}$;\quad ${N}_{cy} = \frac {c_y \Delta t}{h_y}$; 
\quad $Pe_{x} = \frac {\alpha \Delta t}{{h_x}^2}$; \quad $Pe_{y} = \frac {\alpha \Delta t}{{h_y}^2}$. 

For a numerical scheme the corresponding numerical dispersion relation is obtained by analogy as,

\begin{equation}
\omega_{num} = \left(\sqrt{{k_x}^2+{k_y}^2}\right)c_{num}  -  i \alpha_{num}(k_x^2 + k_y^2)
\end{equation}

\noindent where the subscript $num$ denotes that the corresponding property is for the numerical scheme. 

From this numerical dispersion relation, one obtains numerical amplification factor, $G_{num}$, as

\begin{equation}
\displaystyle G_{num} = e^{-i \omega_{num}\Delta t} = e^{-\alpha_{num}({k_x}^2+{k_y}^2) \Delta t} \,\, e^{-i\left(\sqrt{{k_x}^2+{k_y}^2}\right)c_{num}\Delta t}
\label{num_trans_func}
\end{equation}

Employing second order, explicit central difference schemes for all the derivatives in Eq. \eqref{2DLW}, the numerical amplification factor for the 2D LW scheme is 

\begin{equation}
\begin{split}
 G_{num} =& 1-iN_{cx}\sin(k_xh_x)- iN_{cy}\sin(k_yh_y)+(N_{cx}^2+2Pe_x)[\cos(k_xh_x)-1] \\[1.5ex]
         &+(N_{cy}^2+2Pe_y)[\cos(k_yh_y)-1]-N_{cx}N_{cy}\sin(k_xh_x)\sin(k_yh_y)
\end{split}
\label{gn_2dcd2}
\end{equation}

For accuracy, we require $G_{num} /G_{phys} \approx 1$, with $|G_{num}| < 1$. From $G_{num}$, numerical phase speed is obtained as

\begin{equation}
\frac{c_{num}}{c_{phys}} = -\left[\frac{1}{N_{cx}(k_xh_x)+N_{cy}(k_yh_y)}\right]{\tan}^{-1}\left[\frac{(G_{num})_{Img}}{(G_{num})_{Real}}\right]
\label{cnbyc}
\end{equation}

\noindent where $c_{phys}$ is the physical phase speed.

The numerical group velocity components are then obtained as 
\begin{equation}
\frac{v_{gx,num}}{c_x} = \frac{1}{N_{cx}} \frac{\partial\beta_{num}}{\partial(k_xh_x)}
\label{vgnbyc_x}
\end{equation}

\begin{equation}
\frac{v_{gy,num}}{c_y} = \frac{1}{N_{cy}} \frac{\partial\beta_{num}}{\partial(k_yh_y)}
\label{vgnbyc_y}
\end{equation}
\noindent with $\tan(\beta_{num}) = -\left[\frac{(G_{num})_{Img}}{(G_{num})_{Real}}\right]$.

The numerical diffusion coefficient $\alpha_{num}$ is given as
\begin{equation}
\frac{\alpha_{num}}{\alpha} = -\frac{\ln{|G_{num}|}}{[{Pe}_x(k_xh_x)^2+{Pe}_y(k_yh_y)^2]}
\label{num_diff_phy_diff}
\end{equation}

The ratio $\frac{\alpha_{num}}{\alpha}$ determines the numerical diffusion offered by the scheme. If it is unity, then the scheme models the physical diffusion exactly. If the ratio is greater or lesser than unity, then the numerical diffusion is higher or lower than the physical diffusion. Negative value(s) denote anti-diffusion which leads to numerical instability. For accuracy of solution, $\frac{\alpha_{num}}{\alpha}$, $\frac{c_{num}}{c}$, $\frac{v_{gx,num}}{c_x}$ and $\frac{v_{gy,num}}{c_y}$ should be equal to unity. 

An interesting observation can be made by comparing the 2D LW scheme with its 1D counterpart. One notes the presence of an additional cross-derivative term $u_{xy}$ for the 2D case which is absent for the 1D case. From spectral analysis, this cross-derivative term introduces additional numerical dissipation when both $k_x$ and $k_y$ are of the same sign and anti-diffusion when the wavenumbers are of the opposite sign. This behavior of the 2D LW scheme has not been reported before. Hence, it becomes necessary to evaluate the properties of the scheme in order to assess it for simulating fluid flows.

\begin{figure}
        \centering
         \includegraphics[scale=0.55]{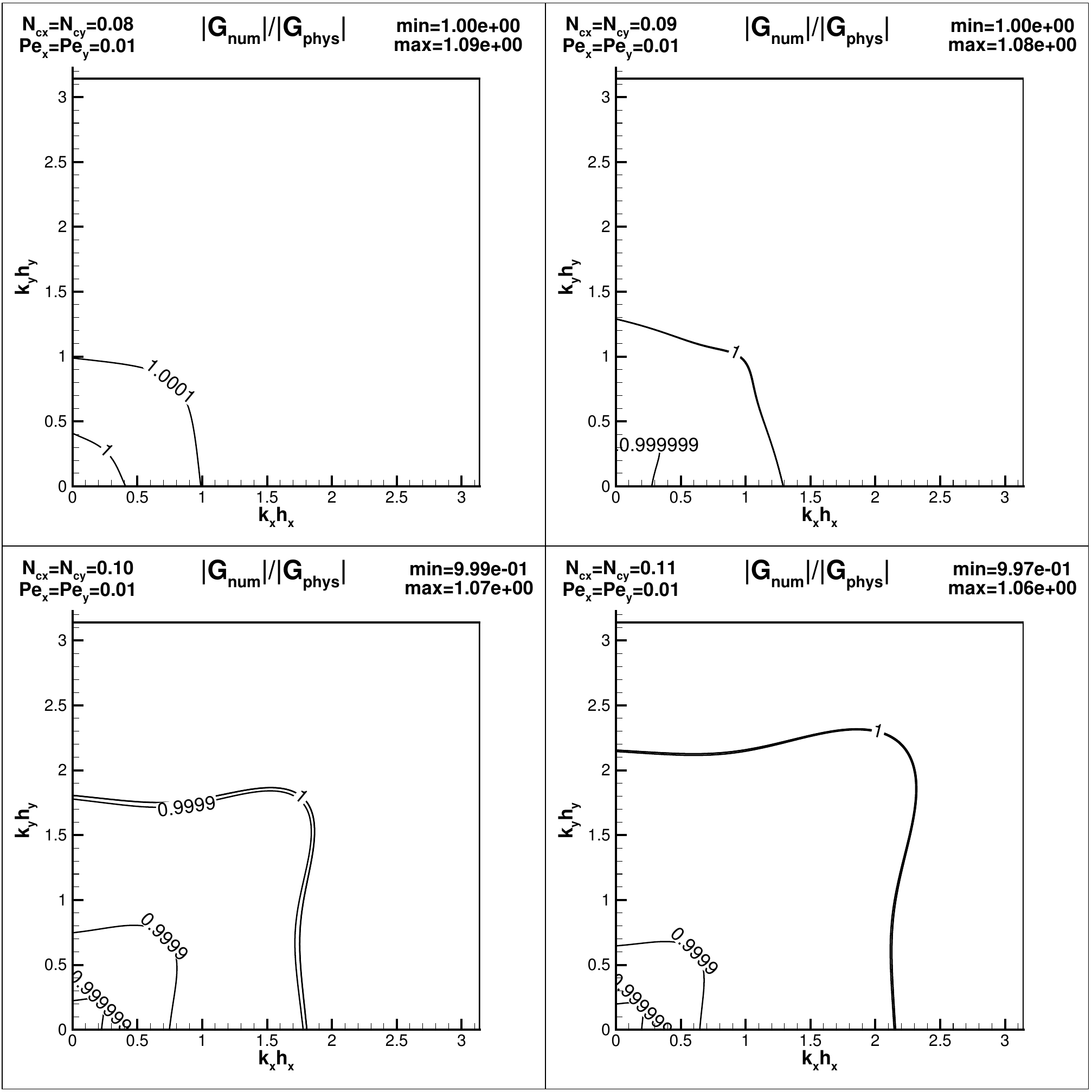}
        \caption{Ratio of numerical to physical amplification factor ($|G_{num}/G_{phys}|$) of the Lax-Wendroff method based on explicit CD stencils for the convection terms in 2D CDE, plotted in the $(k_xh_x,k_yh_y)$-plane for the representative Peclet and CFL numbers. Properties are shown for a uniform grid with equal spacing in both directions and the wave propagating at $45^o$.}
        \label{fig5}
    \end{figure}

In Fig. \ref{fig5}, the ratio of the numerical amplification factor to the physical amplification factor $|G_{num}/G_{phys}|$ is plotted for Peclet numbers $Pe_x=Pe_y=0.01$ and CFL numbers $N_{cx}=N_{cy}=0.08$ to $0.11$. The property charts correspond to the simulation parameters for a uniform grid with equal spacing in $x$ and $y$- directions ($h_x=h_y$) and the solution propagating at an angle of $45^o$ ($\tan(c_y/c_x)=1$). For these simulation parameters, anti-diffusion is absent (not shown in the figure) and hence, the numerical simulations will be stable, i.e. $G_{num} \le 1$. Only two contour values are plotted which correspond to errors of $10^{-4}$ and $10^{-6}$ from the ideal value of $|G_{num}/G_{phys}|=1$. Thus, values of $0.9999$ and $1.0001$ correspond to the tolerance limit of error of $10^{-4}$ while the values $0.999999$ and $1.000001$ denote a lower error of $10^{-6}$, respectively. Comparing the contours for different CFL values, one notes that a maximum extent of region with low error is obtained for $N_{cx}=N_{cy}=0.09$ and hence, this is used as optimal values for the considered $Pe$ values.  

The importance of determining the optimal values for any numerical scheme cannot be underscored as they will help the research community to utilize the scheme in an efficient manner. By identifying optimal $(N_{cx}, N_{cy})$ for given Peclet numbers $(Pe_{x}, Pe_{y})$, timestep $\Delta t$ or grid spacings $h_x, h_y$ can be fixed. One can fully appreciate the vital and critical role played by GSA due to its accurate characterization of numerical schemes for solving specific governing equations. For the present case, it should be noted that only one free variable exists $(N_{cx})$ for determining the optimal conditions as the numerical setup involves an equispaced, uniform grid with signal propagating at $45^o$. 

\begin{figure}
        \centering
         \includegraphics[scale=0.55]{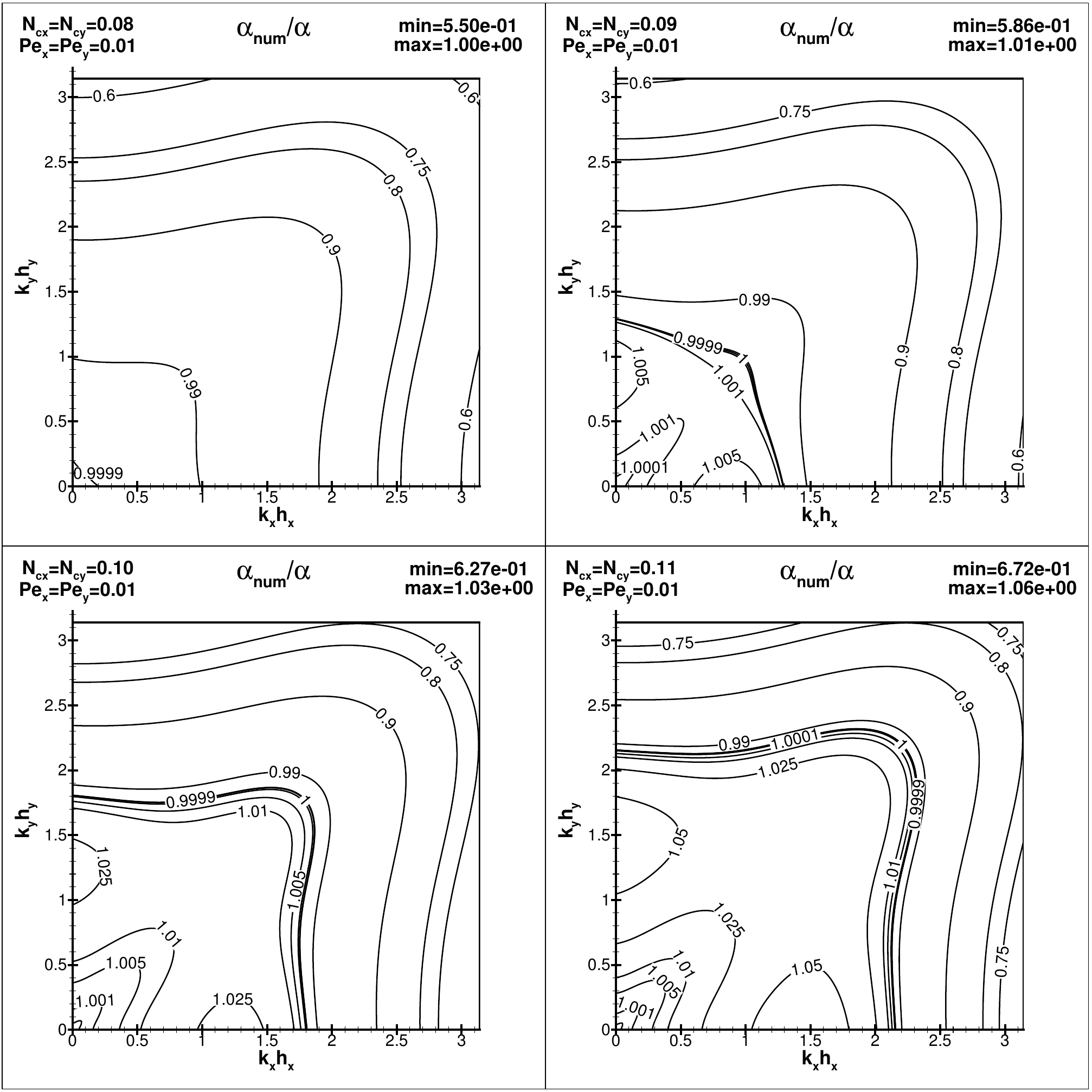}
        \caption{Ratio of numerical to physical diffusion coefficient ($\alpha_{num}/\alpha$) for the Lax-Wendroff method based on explicit CD stencils for the convection terms in 2D CDE, plotted in the $(k_xh_x,k_yh_y)$-plane for the representative Peclet and CFL numbers. Properties are shown for a uniform grid with equal spacing in both directions and the wave propagating at $45^o$.}
        \label{fig_2dalpha}
    \end{figure}

The optimal simulation parameters, evaluated from the analyses based on acceptable errors in $|G_{num}/G_{phys}|$ contours, are reinforced by plotting the ratio of numerical to physical diffusion $\alpha_{num}/\alpha$ for the LW scheme in Fig. \ref{fig_2dalpha} for the same parameters as before. One notes that a maximum resolution for performing fine simulations is obtained for $Pe_x=Pe_y=0.01$ when $N_{cx}=N_{cy}=0.09$. An important observation can be immediately drawn from the figure that the numerical diffusion increases in strength as the CFL value is increased for fixed Peclet values. This is expected as the LW scheme introduces additional numerical dissipation as established in Eq. \eqref{2DLW}. Further, with increase in CFL values, the effectiveness of numerical diffusion increases for higher wavenumbers which can be an advantage in controlling high wavenumber numerical instabilities such as aliasing, arising due to a lack of resolution.   

\begin{figure}
        \centering
         \includegraphics[scale=0.55]{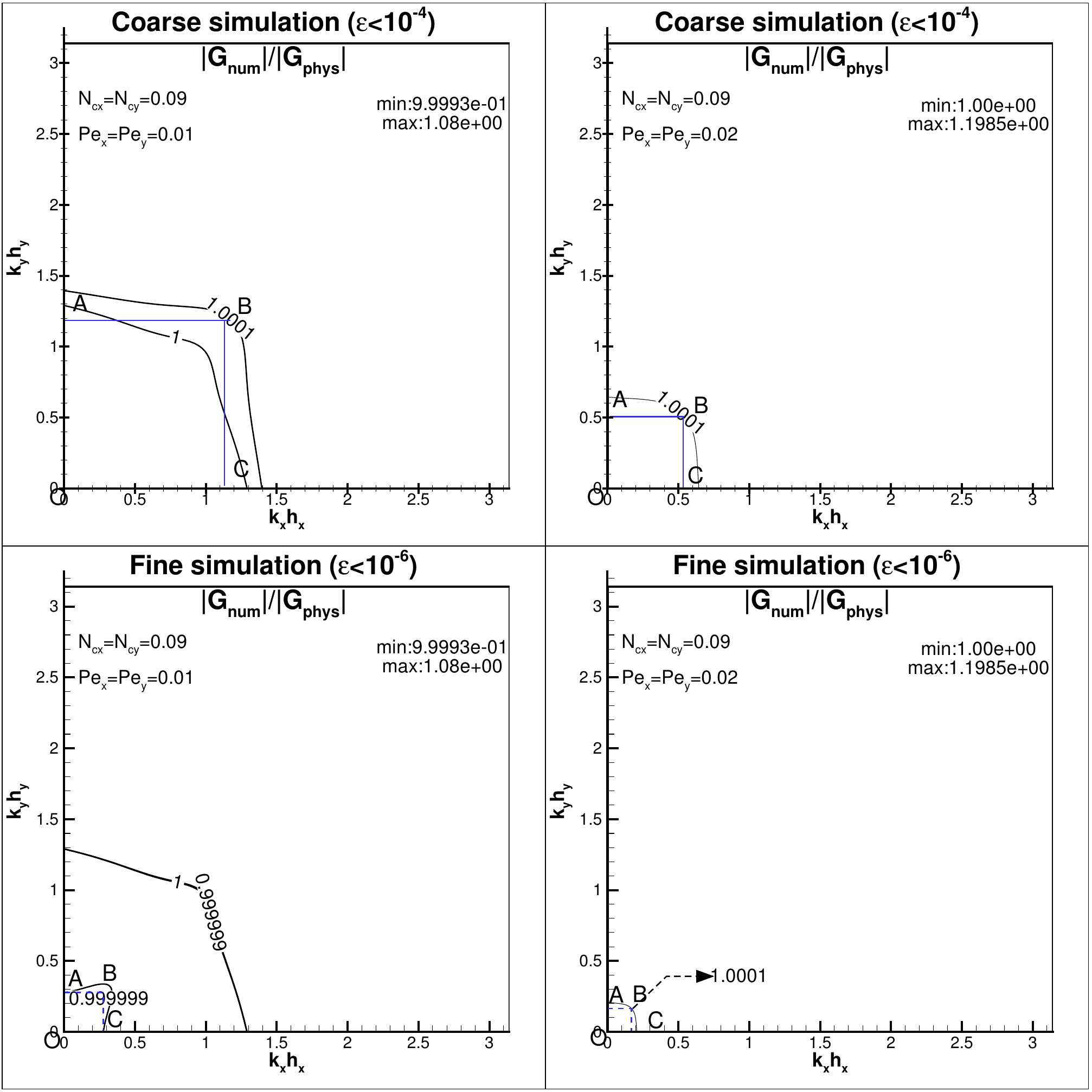}
        \caption{Zones of coarse (top panels) and fine simulations (bottom panels) for the Lax-Wendroff method for explicit CD schemes used for the convection term in 2D CDE, for the indicated Peclet numbers and optimal CFL values of $0.09$. The solid and dashed blue rectangles denote regions with errors $|1-|G_{num}/G_{phys}||  \le 10^{-4}$ and $|1-|G_{num}/G_{phys}|| \le 10^{-6}$, respectively.}
        \label{fig6}
    \end{figure}

In Fig. \ref{fig6}, the regions representing the two tolerance levels are shown for the optimal CFL values for Peclet numbers of $0.01$ and $0.02$, respectively. The boundaries of the regions, denoted by OABC, are chosen so as to maximize the resolutions $k_x$, $k_y$ satisfying the error constraint. The top frames correspond to the higher error tolerances of $10^{-4}$ with the regions marked by solid blue lines in the form of a rectangle and the bottom frames are for the lower error tolerance of $10^{-6}$ with the regions marked by dashed blue rectangles, respectively. In the present work, the former condition is representative of coarse simulations while the latter case denotes fine simulations such as LES/DNS. This is the rationale behind the choice of the two tolerance values. As the tolerance level is reduced, the region satisfying the error criterion shrinks i.e one obtains lower errors by moving towards the continuum limit. It is also interesting to note that the regions for coarse and fine simulations decrease with increasing $Pe$ values. In determining the optimal limits of CFL numbers and the accuracy of the scheme, we have only considered the error in representing the physical amplification factor. However, error is also contributed due to dispersion errors as numerical phase speed $c_{num}$ need not be equal to the physical phase speed $c_{phys}$. 

In the analyses described earlier, we have evaluated the performance of the LW scheme for a small operating range of Pe and $N_c$ values. Therefore, it would only be natural and also essential to quantify the scheme for a wider range of simulation parameters, i.e. determine the accuracy limits of the scheme. It is important to note that in the process of determining these limits, the information on accuracy should also be simultaneously available to the researcher. A direct presentation of this information is a complicated task due to the multidimensional ($\geq3D$) nature of the data. In this regard, we present a simple and a tractable approach that can be followed to achieve the desired goal. First, $\alpha_{num}/\alpha$ property is extracted along a diagonal line $(k_xh_x=k_yh_y)$ for each $N_c$ value and for a fixed Pe number. The resulting data can be then plotted as contours in the $[N_{cx}(=N_{cy}),k_xh_x(=k_yh_y)]$-plane for a fixed Pe value. 

\begin{figure}
        \centering
         \includegraphics[scale=0.55]{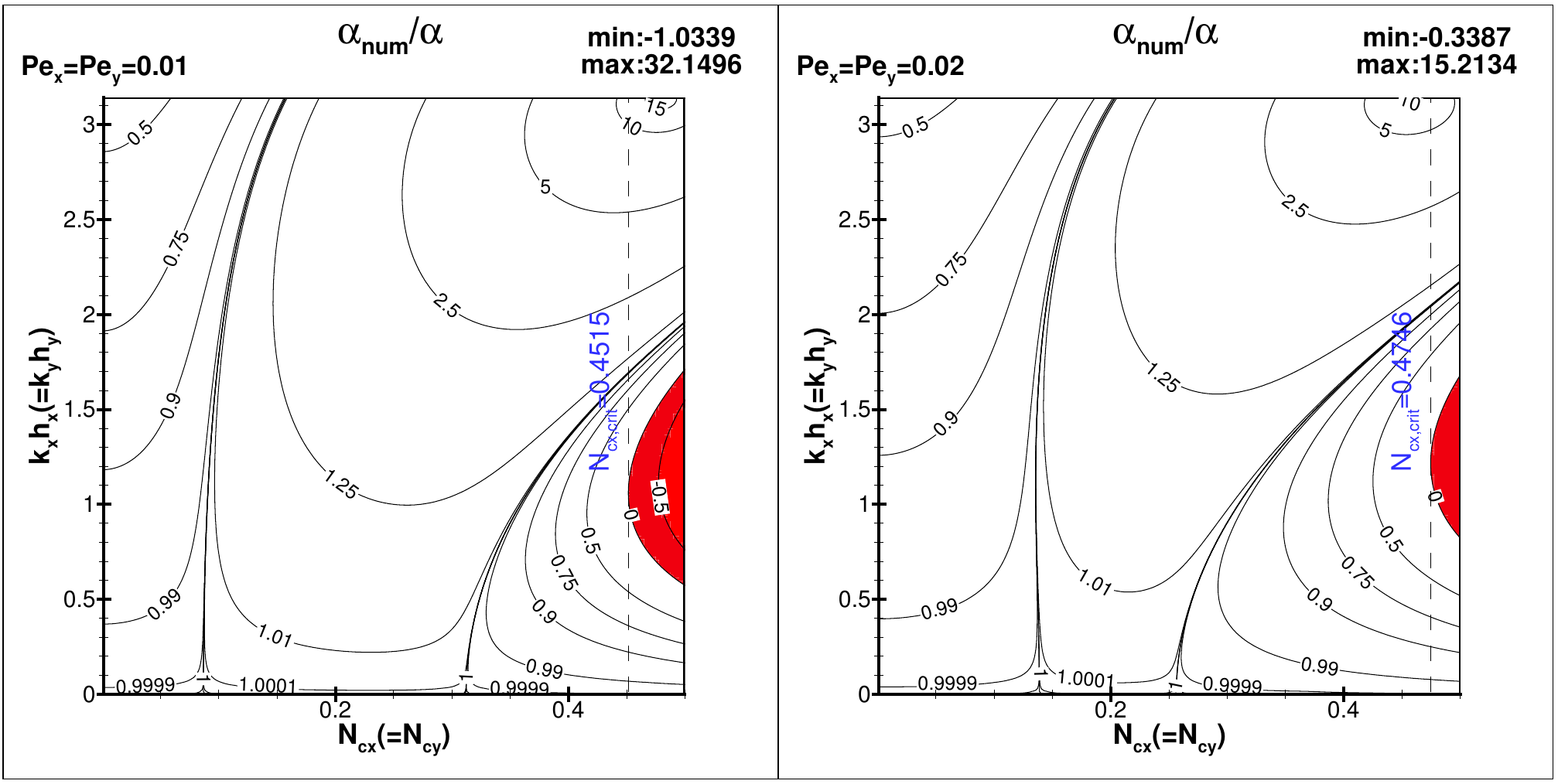}
        \caption{Ratio of numerical to physical diffusion coefficient ($\alpha_{num}/\alpha$) for the Lax-Wendroff method based on explicit CD stencils for the convection terms in 2D CDE, plotted in the $(N_{cx},k_xh_x)$-plane for the representative Peclet numbers. Properties are shown for a uniform grid with equal spacing in both directions and the wave propagating at $45^o$.}
        \label{fig_2dPediag}
    \end{figure}

This approach is demonstrated in Fig. \ref{fig_2dPediag} for the LW scheme for Pe values $0.01$ and $0.02$, respectively. One notes a striking similarity of this figure with the 1D results (Figs. \ref{fig1}, \ref{fig3}) in the previous section. From the results, the stability limits are determined by noting the CFL values ($N_{cx,crit}$) for which anti-diffusion appears. This is noted to be $0.4515$ and $0.4746$ for $Pe_x=0.01$ and $0.02$, respectively. Thus, increasing Pe value has the effect of increasing $N_{cx,crit}$. It is also interesting to note that the results can also be used to determine the optimal $N_{cx}$ values by noting the location(s) at which $\alpha_{num}/\alpha=1$ contour is vertical. This is determined from the figure as $N_{cx}=0.089$ and $0.14$ for $Pe_x=0.01$ and $0.02$, respectively.

Apart from the error of representing the numerical amplification factor accurately, simulations suffer from additional source(s) of error. In \cite{HACM259}, the correct error dynamics equation is derived for the 1D convection equation for the first time using GSA which reveals all the sources of error in computing. Another important contribution of \cite{HACM259} is to dispel the notion that the evolution equation for the error has the same form as the governing equation. In \cite{Bhole}, the governing equation for error evolution is derived for the 1D diffusion equation using GSA showing the contribution to error due to numerical diffusion being different from physical diffusion. Following the same approach, the error dynamics equation for the 2D CDE is presented for the first time as

 \begin{equation}
\begin{split}
e_t + c_xe_x +c_ye_y & = \alpha(e_{xx} +e_{yy})\\ 
&+ \iint\limits_{-k_{max}}^{k_{max}} (\alpha_{num}-\alpha) \; (k_x^2 + k_y^2) \; |G_{num}|^{t^n/\Delta t} \; \hat{U}_0(k_x,k_y) \; e^{i[k_xx+k_yy-(\sqrt{k_x^2+k_y^2})c_{num}t^n]}\,dk_x dk_y \\
&  \iint\limits_{-k_{max}}^{k_{max}} i(\sqrt{k_x^2+k_y^2})(c_{num}-c_{phys}) \; |G_{num}|^{t^n/\Delta t} \; \hat{U}_0(k_x,k_y) \; e^{i[k_xx+k_yy-(\sqrt{k_x^2+k_y^2})c_{num}t^n]}\,dk_x dk_y
\end{split}
\label{2derr}
\end{equation}

Here, $e$ denotes the error and is defined by the difference between exact ($u_{phys}$) and numerical solution ($u_{num}$). From the above equation one notes contribution to error due to incorrect numerical diffusion and phase speed, respectively. The former contribution is also denoted as error due to spurious numerical diffusion/amplification and the latter is the dispersive error. Comparing the error dynamics equation with the MDE for 2D LW method for CDE given by Eq. \eqref{2DLW}, the cross-derivative term is noted to be missing which leads one to believe that the error dynamics equation is incomplete. However, this assertion is incorrect as the missing term is already accounted by $G_{num}$  and $c_{num}$ terms which incorporate the effects of all the terms in MDE including the cross-derivative term. 

The error properties are quantified for the Lax-Wendroff scheme applied to the convection term in the 2D CDE in Tables \ref{tab_quant1} and \ref{tab_quant2} for two Peclet numbers $0.01$ and $0.02$, respectively. These properties are obtained from the rectangular regions OABC identified in Fig. \ref{fig6}. The tables present the maximum resolution, dispersion and dissipation errors exhibited by the numerical scheme. From these results, it is apparent that the maximum resolution satisfying the error constraint is higher for the coarse simulation case when compared to the fine simulation case. We note that the dispersion errors are also maximum for the coarse simulation case. This can be attributed to the poor spectral resolution of the CD$_2$ stencils employed for the Lax-Wendroff scheme. A comparison of the error properties for the two Peclet number cases reveals significant improvement in dispersion and group velocity error properties for the higher Pe value. However, a substantial difference is noted in the maximum resolution for the coarse simulation cases with the higher Pe case showing a reduced resolution. 

\begin{table}[h!]
\centering
\resizebox{\columnwidth}{!}{ \begin{tabular}{|c|c|c|c|c|c|c|c|}
 \hline
Lax-Wendroff& Simulation & Maximum & Maximum & $\alpha_{num}$/ $\alpha$ & $c_{num}$/$c_{phys}$ & $v_{gx,num}$/$c_x$ & $v_{gy,num}$/$c_y$\\
Method for & type & resolution & resolution & range & range & range & range \\ [0.5ex]
& & $(k_xh_x)_{max}$ & $(k_yh_y)_{max}$ & & & & \\ [0.5ex]
 \hline\hline
 Convection term & Coarse & 1.4&1.4& 0.994641-1.00691 & 0.72529-1 & 0.205689-1.02329 & 0.211452-1.02393\\
\hline
 Convection term & Fine& 0.34375&0.34375& 1-1.00125 & 0.95308-1 & 0.95507-1.00081 & 0.953638-1.00082\\[1ex]
 \hline
 \end{tabular}}
 \caption{Quantification of DRP properties for coarse ($|1-|G_{num}/G_{phys}||\le10^{-4}$) and fine simulations ($|1-|G_{num}/G_{phys}||\le10^{-6}$) for 2D CDE using the Lax-Wendroff method based on CD schemes for $Pe_x=Pe_y=0.01$ and $N_{cx}=N_{cy}=0.09$. }
 \label{tab_quant1}
\end{table}

\begin{table}[h!]
\centering
\resizebox{\columnwidth}{!}{ \begin{tabular}{|c|c|c|c|c|c|c|c|}
 \hline
Lax-Wendroff& Simulation & Maximum & Maximum & $\alpha_{num}$/ $\alpha$ & $c_{num}$/$c$ & $v_{gx,num}$/$c_x$ & $v_{gy,num}$/$c_y$\\
Method for & type & resolution & resolution & range & range & range & range \\ [0.5ex]
& & $(k_xh_x)_{max}$ & $(k_yh_y)_{max}$ & & & & \\ [0.5ex]
 \hline\hline
 Convection term & Coarse & 0.64264&0.64264& 0.98811-1 & 0.940531-1 & 0.82691-1.00961 & 0.829639-1.00994\\
\hline
 Convection term & Fine& 0.20498&0.20498& 0.998821-1 & 0.994327-1 & 0.983572-1.00076 & 0.98273-1.0008\\[1ex]
 \hline
 \end{tabular}}
 \caption{Quantification of DRP properties for coarse ($|1-|G_{num}/G_{phys}||\le10^{-4}$) and fine simulations ($|1-|G_{num}/G_{phys}||\le10^{-6}$) for 2D CDE using the Lax-Wendroff method based on CD schemes for $Pe_x=Pe_y=0.02$ and $N_{cx}=N_{cy}=0.09$. }
 \label{tab_quant2}
\end{table}

From the presented property charts and error quantification for the Lax-Wendroff scheme for 2D CDE it can be inferred that accurate computation of Navier-Stokes equation is possible provided one employs timestep as dictated by the optimal CFL values and a finely resolved grid such that all the relevant flow scales are within the maximum resolution dictated by the fine simulation error tolerances. This is demonstrated in the next section by solving the steady and unsteady flow inside a square lid driven cavity (LDC).

\section{LES of 2D NSE by LW Method}
\label{NSEres_sec}
In this section, we demonstrate the utility of the Lax-Wendroff scheme in accurately solving the 2D incompressible Navier-Stokes equation for the flow inside a square LDC for post-critical Reynolds number. This problem is chosen specifically due to its simple, unambiguous boundary conditions and the availability of benchmark solutions for unsteady case \cite{NCCD1,NCCD2}. The post-critical Reynolds number considered in the present study is 10,000. A concise summary of the governing equations and the methodology adopted for the numerical solution is discussed next followed by numerical simulations using the 2D LW scheme.

\subsection{Governing Equations and Solution Methodology}
The governing 2D incompressible NSE is solved in the streamfunction-vorticity ($\psi$-$\omega$) formulation. This formulation is adopted due to its advantages in satisfying mass conservation automatically in the computational domain. In this approach, one solves two equations- a Poisson equation for $\psi$ and a transport equation for $\omega$, which are given below.

\begin{equation}
\nabla^2\psi = -\omega
\label{sfe}
\end{equation}
\begin{equation}
\frac{\partial \omega}{\partial t} + (\vec{V}\cdot\vec{\nabla})\omega = \frac{1}{Re}\nabla^2 \omega
\label{vte}
\end{equation}

These equations are given in their non-dimensional form with $Re$ denoting the reference Reynolds number based on the side of the cavity and the speed of the upper lid which moves from left to right. Velocity vector $\vec{V}=u\hat{i}+v\hat{j}$ is computed from the stream-function by $\vec{V}=\vec{\nabla}\times\vec{\psi}$, where $\vec{\psi} = [ 0\; 0\; \psi ]{^T}$.

The governing equations are solved in the following manner. The stream function equation, Eq. \eqref{sfe} is solved first by using CD$_2$ scheme for discretization and the BiCGSTAB iterative method \cite{BiCG} for eventual solution of the discrete Poisson equation. After this step, the vorticity at the boundaries is computed from its definition: $\omega=-\nabla^2\psi$. Next, vorticity at new time instant is updated by solving Eq. \eqref{vte}, wherein the CD$_2$ Lax-Wendroff scheme applied to convection terms is employed. The Lax-Wendroff scheme for the 2D VTE is given in Eq. \eqref{2dnse_lw} for reference. The process is repeated until the simulation time reaches a maximum user specified value or when the flow reaches a steady state. In performing these steps, one should also note that a constant value, Dirichlet boundary condition is imposed on $\psi$ ($\psi=\psi_0$) in order to satisfy the non-penetrative condition at the solid walls.

\begin{equation}
\begin{split}
\omega_{ij}^{n+1} = \omega_{ij}^{n} + \Delta t \left(-u\frac{\partial \omega}{\partial x}-v\frac{\partial \omega}{\partial y}\right)_{ij}^{n} + \frac{\Delta t^2}{2}&\left[\left(u\frac{\partial u}{\partial x}+v\frac{\partial u}{\partial y}\right)\frac{\partial \omega}{\partial x} + u^2 \frac{\partial^2 \omega}{\partial x^2} + \left(u\frac{\partial v}{\partial x}+v\frac{\partial v}{\partial y}\right)\frac{\partial \omega}{\partial y}\right.\\[1.5ex]
&\left. +2uv\frac{\partial^2 \omega}{\partial x\partial y} + v^2\frac{\partial^2 \omega}{\partial y^2}\right]_{ij}^{n} + \frac{1}{Re} \nabla^2 \omega_{ij}^{n}
\end{split}
\label{2dnse_lw}
\end{equation}

\subsection{Numerical Solution at LDC at Post-critical Re=10,000}
The LDC problem is solved here for a post-critical Re of 10,000 using the explicit CD based Lax-Wendroff scheme. The present case is chosen as it is an excellent test case for benchmarking the accuracy of numerical schemes/codes. Accurate solutions for this Re must display transient, triangular vortex structure during the flow evolution \cite{NCCD1,NCCD2}. Hence, its capture serves as a direct confirmation of accuracy for unsteady flows. The numerical solution is obtained for a uniform, equispaced grid of $1112\times1112$ with a timestep of $\Delta t=8.1\times10^{-5}$. These simulation parameters result in $Pe_x=Pe_y=0.01$ and CFL number based on upper lid velocity $N_{cx}=0.09$, respectively and correspond to the optimal conditions determined for the scheme. 

\begin{figure}
        \centering
         \includegraphics[height=0.9\textheight,keepaspectratio=true]{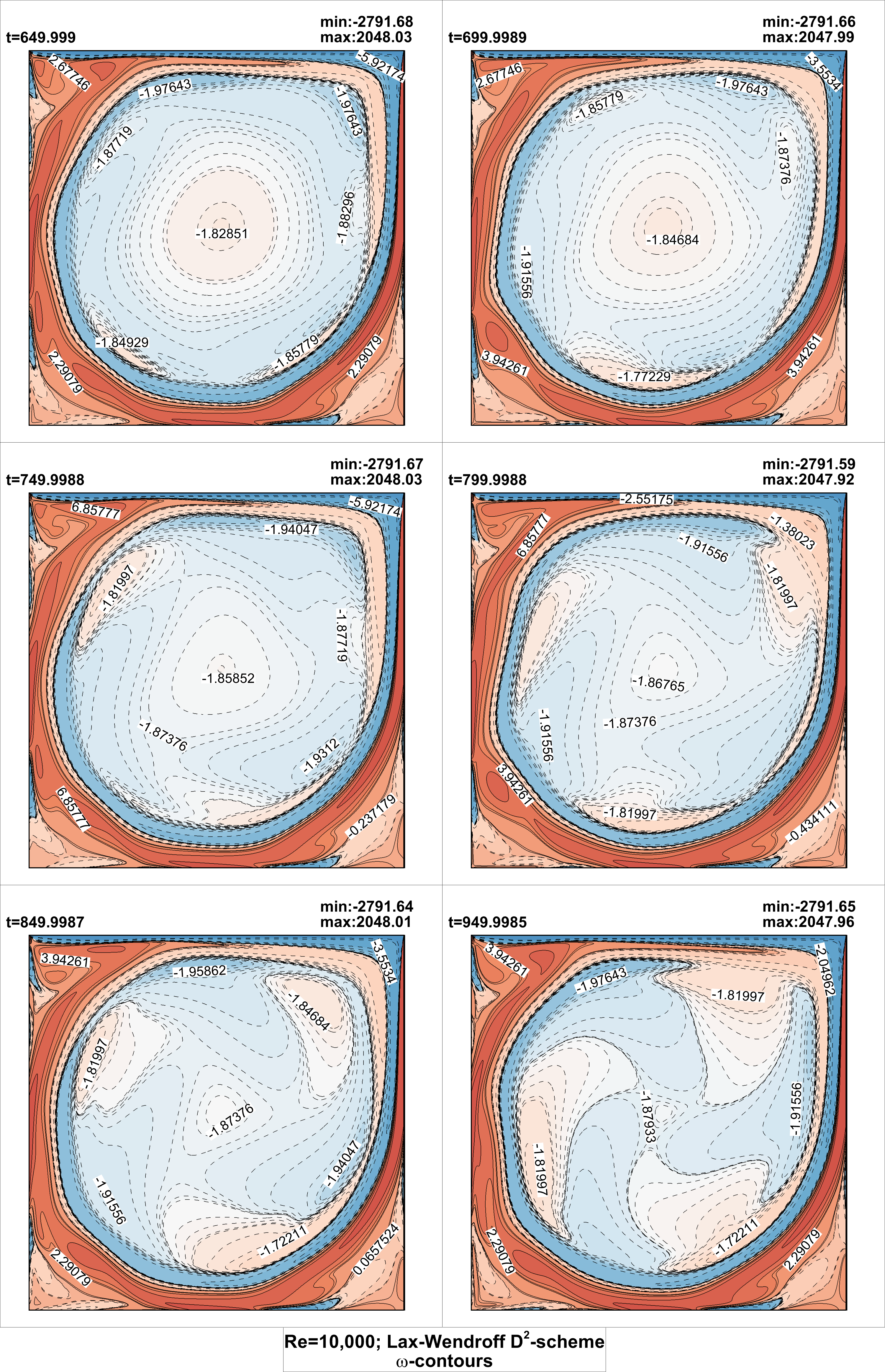}
        \caption{Evolution of vorticity field showing the transient, triangular core vortex for the LDC problem for Re=10,000 computed by the explicit central difference based Lax-Wendroff method applied to convection terms of 2D incompressible NSE.}
        \label{fig9}
    \end{figure}

In Fig. \ref{fig9}, vorticity contours are plotted at the indicated time instants showing the complex vortex dynamics of the flow. At early times, one notes the formation of a pentagonal vortical structure in the core which subsequently evolves into a triangular structure due to the shearing action. The triangular vortex is observed which is surrounded by gyrating satellites which rotate along with it. The triangular vortex structure shrinks in size at much later times thus displaying its transient nature. This evolution provides an excellent match with the observations in \cite{NCCD1,NCCD2} which was used for high accuracy combined compact schemes. The present results, therefore, demonstrates the potential of the Lax-Wendroff scheme for solving unsteady flows and further consolidates the utility of GSA in determining the optimal limits for simulation parameters for realizing a desired level of accuracy.

\section{Summary and Conclusions}
\label{sumcon_sec}
In the present paper, the Lax-Wendroff (LW) method based on explicit central differences is comprehensively analyzed using global spectral analysis (GSA) for the model 1D and 2D linear convection-diffusion equations (CDE) in order to obtain optimal simulation parameters for performing accurate implicit LES (ILES). The LW method is one of the earliest approaches in developing second order in time DRP schemes via the modified differential equation (MDE), where higher order time derivatives are converted to spatial derivatives using the governing equation. While the LW scheme was analyzed recently using GSA, important questions regarding its variants, analysis for higher than one dimensional problems and optimal parameters for accuracy are addressed here in brief. The optimal simulation parameters presented here ensure accurate simulation of CDE and thereby Navier-Stokes equations, ILES to be specific. This is achieved by a rigorous quantification of the numerical properties viz. the resolution, numerical diffusion, dispersion and signal propagation speed for the LW method for 2D CDE. 

A consequence of the application of the GSA for the 1D and 2D CDE is that the diffusion and solution propagation properties for numerical schemes become space-time dependent. Hence, the numerical schemes have corresponding wavenumber dependent coefficients such as numerical diffusion coefficient $\alpha_{num}$, numerical phase speed $c_{num}$ and numerical group velocity $v_{g,num}$, which determine the evolution of the solution. To accurately solve the governing equations, these properties must match their physical counterparts, and thereby these define dispersion relation preserving (DRP) schemes.

Formulating the MDE for the 1D CDE, two variants of the LW method are developed based on the treatment of convection and convection-diffusion terms i) as applied only to the convection term (3$^{rd}$ and 4$^{th}$ derivatives are set to zero in Eq. \eqref{eqn25}) and ii) applied to all the terms given in Eq. \eqref{eqn25}, respectively. The first approach is popularized and practiced by many practitioners without adequate numerical analysis, while the second approach is the full LW scheme in the present context. One of the interesting observations noted in the application of the full LW procedure for 1D CDE is the inclusion of a third and fourth order derivative terms along with an added second order derivative term. The latter is noted as the reason for the LW method to be inconsistent for the 1D convection equation. Presence of the third and fourth derivatives introduce the effect of dispersion and weak anti-diffusion when using the LW method with the latter moderating the added numerical diffusion to the already existing physical term.

GSA shows marginal benefits offered by the full LW scheme over the other variant. Analysis of two values of Peclet numbers $0.01$ and $0.02$, shows the full LW method to possess an increased stability region over the other variant. The analysis also reveals parameter combinations in the $(N_c,kh)$-plane for both schemes where the overall numerical diffusion is equal to the physical diffusion despite the added second and fourth order numerical diffusion terms. The Lax-Wendroff method applied to convection term alone shows better performance because of the near vertical contour line ${\alpha_{num}\over\alpha} =1$, offering more resolution over the full scheme. Furthermore, the full LW scheme is computationally more expensive due to the $3^{rd}$ and $4^{th}$ order derivative terms. This provides the necessary justification for adopting the LW method based on its application to the convection term. Hence, the LW scheme applied to convection only terms is analyzed for 2D CDE in order to determine optimal parameters for ILES and Navier-Stokes simulations.

The LW method is evaluated for solution of 2D CDE using GSA due to a one-to-one correspondence between the model equation and the Navier-Stokes equations \cite{B118}. An additional cross-derivative term is noted for the scheme which is reported for the first time and whose effect is to cause asymmetry in the numerical properties. Rigorous quantification/ evaluation of the DRP region is performed to determine the best numerical parameters viz. a time step/ grid resolution for accurate ILES. This is evaluated by prescribing tolerance limits of $10^{-4}$ and $10^{-6}$ in $|1-|{G_{num} \over G_{phys}}||$ i.e. $|1-|{G_{num} \over G_{phys}}|| \le \epsilon$, with $\epsilon$ as the tolerance. The latter tolerance limit can be considered as representative of ILES and unresolved DNS scenarios and such cases are termed as fine simulations. Optimal CFL values are determined for representative Pe values for which the scheme performs the best as noted from Figs. \ref{fig5} and \ref{fig_2dalpha}. In determining these limits, only the error in numerical amplification factor is considered. Analysis of the complete sources of error for the 2D CDE given by Eq. \eqref{2derr} also shows contribution from the dispersion property. A complete assessment of the performance of the method with respect to these errors are presented in Tables \ref{tab_quant1} and \ref{tab_quant2}. It is also noted that as CFL or Peclet number increases, the scheme becomes more stable.

The optimal parameters obtained for the LW scheme are corroborated by solving 2D Navier-Stokes equations for the square lid driven cavity problem for post-critical Reynolds number of $10,000$. For the post-critical case optimal CFL conditions are used as the time evolution is important. The simulation shows a very good agreement with the benchmark results for the unsteady case capturing the transient triangular vortex which is considered as a stringent test for validating high accuracy solution methods. The present work demonstrates accurate analyses of the LW method as corroborated by Navier-Stokes simulation thus highlighting the potential of GSA approach. 

\section*{Acknowledgments}

\subsection*{Author contributions}
All authors have contributed equally to the data generation, analysis, writing and preparation of the manuscript.

\subsection*{Financial disclosure}

None reported.

\subsection*{Conflict of interest}

The authors declare no potential conflict of interests.

\section*{Supporting information}
None available.
 
\nocite{*}

\end{document}